\documentstyle[aps,eqsecnum,epsfig,preprint]{revtex}\tighten

\def\vec{\bf} 

\begin{document}
\title{Real-time nonequilibrium dynamics in hot QED plasmas:\\
dynamical renormalization group approach}
\author{\bf S.-Y. Wang,$^{1}$ D. Boyanovsky,$^{1,2}$ H. J. de
Vega,$^{2,1}$ and D.-S. Lee$^{3}$}
\address
{${}^{1}$Department of Physics and Astronomy, University of
Pittsburgh, Pittsburgh,  Pennsylvania 15260, USA\\
${}^{2}$LPTHE, Universit\'e Pierre et Marie Curie (Paris VI) et Denis Diderot
(Paris VII), Tour 16, 1er. \'etage, 4, Place Jussieu, 75252 Paris,
Cedex 05, France\\
${}^{3}$Department of Physics, National Dong Hwa University, Shoufeng,
Hualien 974, Taiwan, Republic of China}
\date{May 2000}
\maketitle

\begin{abstract}
We study the real-time nonequilibrium dynamics in hot QED plasmas
implementing a dynamical renormalization group and using the hard
thermal loop (HTL) approximation. The focus is on the study of the
relaxation of gauge and fermionic mean fields and on the quantum
kinetics of the photon and fermion distribution functions. For
semihard photons of momentum $eT\ll k \ll T$ we find to leading
order in the HTL that the gauge mean field relaxes in time with a
power law as a result of infrared enhancement of the spectral
density near the Landau damping threshold. The distribution
function of semihard photons in linear response also relaxes with
a power law, with a power that is twice that for the mean field.
The dynamical renormalization group reveals the emergence of
detailed balance for microscopic time scales larger than $1/k$
while the rates are still {\em varying} with time. The quantum
kinetic equation for the photon distribution function allows us to
study photon production from a thermalized quark-gluon plasma
(QGP) by off-shell effects. We find that for a QGP of temperature
$T \sim 200\;\mbox{MeV}$ and lifetime $10\lesssim t \lesssim
50\;\mbox{fm}/c$ the hard ($k \sim T$) photon production from
off-shell bremsstrahlung ($q\rightarrow q\gamma$ and
$\bar{q}\rightarrow\bar{q}\gamma$) at ${\cal O}(\alpha)$ grows
logarithmically in time and is {\em comparable} to that produced
from on-shell Compton scattering and pair annihilation at ${\cal
O}(\alpha\,\alpha_s)$. Hard fermion mean fields relax as
$e^{-\alpha\,T\,t\,\ln(\omega_P t)}$ with $\omega_P=eT/3$ the
plasma frequency, as a consequence of the emission and absorption
of soft magnetic photons. A quantum kinetic equation for hard
fermions is obtained directly in real time from a field
theoretical approach improved by the dynamical renormalization
group. The collision kernel is {\em time-dependent} and {\em
infrared finite}. In linear response the fermion distribution
function relaxes with an anomalous exponential law with an
exponent twice as large as that for the mean field.
\end{abstract}


\section{Introduction}

The study of nonequilibrium phenomena under extreme conditions
play a fundamental role in the understanding of ultrarelativistic
heavy ion collisions and early universe cosmology. Forthcoming
relativistic heavy ion experiments at BNL Relativistic Heavy Ion
Collider (RHIC) and CERN Large Hadron Collider (LHC) aim to search
for a deconfined  phase of quarks and gluons, the quark gluon
plasma (QGP), which is  predicted by lattice QCD
simulations~\cite{qgp1,qgp2} to emerge at a temperature scale
$T\gtrsim 200\;\mbox{MeV}$.

Recent results from CERN Super Proton Synchrotron
(SPS)~\cite{cern} seem to confirm the main theoretical ideas that
in the central region in ultrarelativistic heavy ion collision a
deconfined plasma of quarks and gluons forms which expands and
cools rapidly and eventually hadronizes.

Current estimates based on energy deposited in the central region
for $\sqrt{s}/\text{nucleon-pair}\sim 200\;{\rm GeV}$ at RHIC
suggest that the lifetime of a deconfined QGP is of order  $10-50~
\mbox{fm}/c$~\cite{qgp1,qgp2}. At such unprecedented short time
scales, an important aspect is an assessment of thermalization
time scales and the potential for non-equilibrium effects
associated with the rapid expansion and finite lifetime of the
plasma and their impact on experimental observables. Lattice QCD
is simply unable to deal with these questions because simulations
are restricted to thermodynamic equilibrium quantities and a
field-theoretical nonequilibrium approach is needed for an
accurate description of the formation and evolution of the QGP. An
important and pioneering step in this direction was undertaken by
Geiger~\cite{geiger} who applied transport methods combined with
perturbative QCD (pQCD) to obtain a quantitative picture of the
evolution of partons in the early stages of formation and
evolution of the plasma.

The consistent study of the evolution of partons in terms of pQCD
cross sections that include screening corrections to avoid the
infrared divergences associated with small angle scattering lead
to the conclusion that quarks and gluons thermalize on time scales
of a few $\mbox{fm}/c$~\cite{thermalization}.  The necessity of a
deeper understanding of equilibrium and non-equilibrium aspects of
the quark-gluon plasma motivated an intense study of the abelian
and non-abelian plasmas in extreme environments. A major step
towards a consistent description of non-perturbative aspects was
taken by Braaten and Pisarski~\cite{htl,rob2,rob3,weldon1,robinfra},
who introduced a novel resummation method that re-organizes the perturbative
expansion in terms of the degrees of freedom associated with
collective modes~\cite{weldon1}, rather than bare particles. This
program, called the hard thermal loop or HTL program is now at the
heart of most treatments of equilibrium aspects of abelian and
non-abelian plasmas~\cite{lebellac}.

Thermal field theory provides the tools to study the properties of
plasmas in {\em equilibrium}~\cite{lebellac,inmedium,yaffebrown,plasmas},
but the consistent study of nonequilibrium phenomena in real time requires
the methods of nonequilibrium field theory~\cite{noneq} (see
also Refs.~\cite{disip,tadpole} for further references). The study of the
equilibrium and nonequilibrium properties of abelian and non-abelian
plasmas as applied to the QGP has as ultimate goal a deeper understanding
of the potential experimental signatures of the formation and evolution of
the QGP in ultrarelativistic heavy ion collisions.
Amongst these, photons and dileptons (electron and/or muon pairs)
produced during the early stages of the QGP  are considered as some of
the most promising signals~\cite{photdilep,kapusta,baier}. Since
photons and lepton pairs interact electromagnetically their mean free paths are
longer than the estimated size of the QGP fireball $\sim 10-50\;
\mbox{fm}$ and unlike hadronic signals they do not undergo final state
interactions. Therefore photons and  dileptons produced during the
early stages of QGP carry clean information from this phase.

\vspace{3mm}

{ \bf The goals of this work}.
In this work we aim to provide a comprehensive study of several
relevant aspects of the {\em nonequilibrium} dynamics of an abelian
QED plasma in {\em real time}. Many features of full QCD are similar
to those of the abelian (QED) theory\cite{htl,lebellac,weldon1}, in
particular leading contributions in the HTL limit can be straighfordarly
generalized from QED to QCD, thus the leading results for
relaxation and photon  production from a QGP\cite{kapusta,baier} can
be understood from the study of a QED plasma.

We utilize a gauge invariant formulation,
available in abelian gauge theories that circumvents the possible
ambiguities associated with gauge invariance~\cite{gaugepot,plasmino}.
Thus gauge and fermionic mean fields and distribution functions are
automatically gauge invariant. We implement and apply the method of
the dynamical renormalization
group, introduced recently to study non-equilibrium phenomena directly in
real time~\cite{boyanrgir,boyanrgk} to extract a consistent,
non-perturbative description of real-time dynamics out of equilibrium.

In particular we focus on the following:

\begin{itemize}
\item[(i)]{The real time evolution of gauge mean fields in linear
response in the HTL approximation. The goal here is to study
directly in real time the relaxation of (coherent) gauge field configurations
in the linearized approximation to leading
order in the HTL program. Whereas a similar study has been carried out in
{\em scalar} QED~\cite{boyanhtl} and confirmed numerically in~\cite{rajantie}
the most relevant case of spinor QED has not yet been studied in detail.}

\item[(ii)]{The quantum kinetic equation that describes the evolution of the
distribution function of photons in the medium, again to leading order
in the HTL approximation. This aspect is relevant to study photon
production via off-shell effects directly in real time. As explained
in detail, this quantum kinetic equation, obtained from a microscopic
field theoretical approach based on the dynamical renormalization
group~\cite{boyanrgir,boyanrgk} displays novel off-shell effects that
cannot be captured via the usual kinetic description that assumes
completed collisions~\cite{photdilep,kapusta,baier}.}

\item[(iii)]{The evolution in real time of fermionic mean fields
features anomalous relaxation arising from the emission and
absorption of magnetic photons (gluons) which are only dynamically
screened by Landau damping~\cite{robinfra,blaizotBN,taka}.
The fermion propagator was studied previously in real time in the
Bloch-Nordsieck approximation which provides a resummation of the
infrared divergences associated with soft photon (or gluon)
bremsstrahlung in the medium~\cite{blaizotBN,taka}. In this
article we implement the dynamical renormalization group to study the
evolution of fermionic mean fields providing an  alternative to the
Bloch-Nordsieck treatment.}

\item[(iv)]{We obtain the quantum kinetic equation for the fermionic
distribution function for hard fermions via the implementation of the
dynamical renormalization group. There has recently been an important
effort in trying to obtain the effective kinetic (Boltzmann) equations
for hard charged
(quasi)particles~\cite{bodeker,manuel,BlaizotBOLT} but the
collision kernel in this equation features the logarithmic divergences
associated with the emission and absorption of
soft magnetic photons (or gluons)~\cite{BlaizotBOLT}. The dynamical
renormalization group leads to a quantum kinetic equation
directly in real time bypassing the assumption of completed collisions
and leads to a {\em time-dependent} collision
kernel free of infrared divergences.}
\end{itemize}

{ \bf Summary of the main results}. The main results of this
study are summarized as follows.

\begin{itemize}
\item[]{{\bf Relaxation of gauge mean fields}. We studied the
relaxation of a gauge mean field in linear response to leading
order in the HTL approximation both for soft momentum $ k \lesssim  e T$ and
for semihard momentum  $ eT \ll k \ll T $ under the assumption of
weak electromagnetic coupling.

{\em Soft momentum} ($k \sim eT$): in this case
 the relaxation of the gauge mean field is
dominated by the end-point contribution of the Landau damping cut.
As a consequence, the soft gauge mean field relaxes with a power law
long time tail of the form
$$
{\vec a}_T({\vec k},t) \buildrel{kt\gg 1}\over=
{\bf a}_T({\vec k},0)
\left[\frac{k^2 Z_T(k)}{\omega^2_T(k)}\cos[\omega_T(k)t]
-\frac{12}{e^2 T^2}\frac{\cos kt}{t^2}\right]\; ,
$$
where $\omega_T(k)$ is the transverse photon pole and $Z_T(k)$
is the corresponding residue.
We note that in spite of the power law tail the gauge mean
field relaxes towards the oscillatory mode determined by the
transverse photon pole. This reveals that the soft collective excitation
in a plasma is stable in the HTL approximation.

{\em Ultrasoft momentum} ($ k \ll  eT$): In the region of ultrasoft momentum
the spectral density divided by the frequency  features a sharp
Breit-Wigner peak near zero
frequency in the region of Landau damping, with width $\Gamma_k =
12k^3/\pi e^2 T^2 $. We find that
the amplitude of a mean field of transverse photons prepared via a source that
is adiabatically switched-on is given by
$$
{\bf a}_T({\vec k},t)\buildrel{kt\gg 1\, , \, \Gamma_k \, t \lesssim
1}\over= {\bf a}_T({\vec k},0)
\left[\frac{k^2 Z_T(k)}{\omega^2_T(k)}\cos[\omega_T(k)t]+
e^{-\Gamma_k t}\right]\; .
$$
We emphasize that the exponential decay is a consequence of the
sharp resonance near zero frequency in
the Landau damping region of the spectral density and
{\em only} arises if the mean field is prepared by an external source
whose time Fourier transform has a simple pole at
zero frequency, such is the case for an adiabatically prepared initial
state. This result confirms those found numerically in Ref.~\cite{rajantie}.

{\em Semi-hard momentum} ($ eT \ll k \ll T $):
In this region both the HTL approximation and the perturbative
expansion are formally valid.
However the spectral density in the Landau damping region is sharply peaked near
$\omega = k$ and the transverse photon pole approaches the
edge of the Landau damping region from above. We find that although
the perturbative expansion is in principle valid, the
sharp spectral density near the edge of the continuum results in a
breakdown of the perturbative expansion. The dynamical
renormalization group provides a consistent resummation  of the lowest order
HTL perturbative contributions in real
time, leading to the following relaxation of the mean field at intermediate
asymptotic times:
$$
{\bf a}_T({\vec k},t) \buildrel{kt \gg 1}\over= \bbox{\cal A}_T({\vec
k},t) \left( \frac{t}{\tau_0} \right)^{-\frac{e^2T^2}{12k^2}}\;,
$$
where $\tau_0\sim 1/k$ and $\bbox{\cal A}_T({\vec k},t)$ is an
oscillating function.
The anomalous exponent is a consequence of an infrared enhancement
arising from the sharp spectral density near the threshold of the
Landau damping region for semihard momentum. The crossover to
exponential relaxation due to collisional processes at higher orders
is discussed.}

\item[]{{\bf Quantum kinetic equation for the photon distribution function}.
Using the techniques of nonequilibrium
field theory and the dynamical renormalization group, we obtain the
quantum kinetic equation for the distribution
function of semihard photons $eT\ll k \ll T$ to lowest order in the HTL
approximation assuming that the fermions are thermalized. An
important result is that the collision kernel is {\em time-dependent}
and the dynamical renormalization group reveals that detailed balance
emerges during microscopic time scales, i.e,
much shorter than the relaxation scales. In the linearized
approximation we find that the departure from equilibrium
of the photon distribution function relaxes as:
$$
\delta n^\gamma_{\vec k}(t)=\delta n^\gamma_{\vec k}(t_0)
\left(\frac{t-t_0}{\tau_0}\right)^{-\frac{e^2 T^2}{6k^2}}
\quad\mbox{for}\quad k(t-t_0)\gg 1\; ,
$$
where $\tau_0\sim 1/k$, and $t_0$ is the initial time.
Furthermore, this quantum kinetic equation
allows us to study photon production by off-shell
effects, which to leading order in the HTL approximation are
determined by photon bremsstrahlung and is of order $\alpha$.
Extrapolating the result from QED to thermalized QGP with two
flavors of light quarks, we find that the total number of hard
photons at time $t$ per invariant phase space volume to lowest order is
$$
N(t) =  \frac{5  \alpha T^3}{18 \pi^2 k^2} \left\{\ln \left[
2k(t-t_0)\right]+\gamma_E-1\right\} \quad \mbox{for} \quad k(t-t_0) > 1\;.
$$
with $t_0 \approx 1\mbox{fm/c}$ is the time scale at which the QGP plasma is thermalized.

We find that for a quark-gluon plasma at temperature $T \sim 200\;\mbox{MeV}$
and of lifetime $10 \lesssim (t-t_0) \lesssim 50\;\mbox{fm}/c$, the
hard ($k\sim T$)
photon production by off-shell bremsstrahlung is comparable to
that from Compton scattering and pair annihilation of
order $\alpha\,\alpha_s$~\cite{kapusta,baier}.}

\item[]{{\bf Relaxation of fermion mean fields}.
We implement the dynamical renormalization group resummation to study the
real-time relaxation of a fermion mean field for hard momentum. The
emission and absorption of magnetic photons which
are only dynamically screened by Landau damping introduce a logarithmic
divergence in the spectral density near the fermion mass shell.
The dynamical renormalization group resums these
divergences in real time and leads to a relaxation of the
fermion mean field for hard momentum given by:
$$
\psi({\vec k},t)\buildrel{kt \gg 1}\over= e^{-\alpha
Tt\left[\ln(\omega_P t)+0.12652\ldots \right]} \times
\mbox{oscillating phases},
$$
with $\omega_P$ being the plasma frequency.}

\item[]{{\bf Quantum kinetics for the fermion distribution function}.
We obtain a quantum kinetic equation for the
distribution function of hard fermions using non-equilibrium field
theory and the dynamical renormalization group resummation. Assuming
thermalized photons we find that the collision kernel
is {\em infrared finite but time-dependent}.
In the linearized approximation the distribution
function relaxes as:
$$
\delta n^f_{\vec k}(t)\buildrel{kt \gg 1}\over=
\delta n^f_{\vec k}(t_0)\; e^{-2\alpha T(t-t_0)\,
\left[\ln(\omega_P t)+0.12652\ldots\right]}\;,
$$
where the anomalous relaxation exponent is twice that of the mean field.}
\end{itemize}

The article is organized as follows.
In Sec.~II we review the main ingredients of nonequilibrium field theory,
the initial value problem formulation for relaxation of nonequilibrium
mean fields, and the dynamical renormalization group approach to quantum kinetics.
In Sec.~III we first study relaxation of the photon mean field in the
hard thermal loop limit for soft $k\leq eT$ and  semihard $eT\ll k
\ll T$ photon momentum and obtain the quantum kinetic equation for the
(hard and semihard) photon
distribution function. In Sec.~IV we study relaxation of fermionic
mean fields for hard momentum and the quantum kinetic
equation for the fermion distribution function.
Our conclusions and some further questions are presented in Sec.~V.

\section{General Aspects}

As stated in the introduction our goal is to provide a systematic
study of non-equilibrium phenomena
in a hot abelian plasma. In particular we focus on a detailed study of
the real time relaxation of expectation values
of fermions and gauge bosons, i.e, fermionic and gauge mean fields as
well as the quantum kinetics for the evolution
of the expectation value of the number operator associated with
fermions and photons. A fundamental issue that must
be addressed prior to setting up our study is that of gauge
invariance. In the abelian case it is straightforward to
reduce the Hilbert space to the gauge invariant subspace and to define
gauge invariant charged fermionic operators. The
description in terms of gauge invariant states and operators is best
achieved within the canonical formulation which
begins with the identification of the canonical field and conjugate
momenta and the primary and secondary first class constraints
associated with gauge invariance. The physical states are those
annihilated by the constraints and physical
operators commute with the (first class) constraints. This program has
been implemented explicitly in the case of scalar quantum
electrodynamics~\cite{boyanrgk,gaugepot} and the fermionic case can be
treated in the same manner with few  minor technical modifications.

The final result of this formulation is that the Hamiltonian acting on
the gauge invariant states and written in terms of
gauge invariant fields is exactly equivalent to that obtained in
Coulomb gauge, which is the statement that Coulomb
gauge describes the theory in terms of the physical degrees of
freedom.  Furthermore the instantaneous Coulomb interaction can be
traded for a Lagrange multiplier~\cite{boyanrgk} leading
to the following Lagrangian density
\begin{eqnarray*}
{\cal L}&=& \bar{\Psi}\left(i{\not\!{\partial}}-e\gamma_0 A_0+e
{\bbox \gamma}\cdot{\vec A}_T-m\right)\Psi
+\frac{1}{2}\left[\left(\partial_\mu {\vec A}_T\right)^2
+ \left(\nabla A_0\right)^2\right]\;.
\end{eqnarray*}
\noindent where ${\vec A}_T$ is the transverse component of the vector
potential and $A_0$ is {\em not} to be interpreted as the time
component of the gauge vector potential but is the
Lagrange multiplier associated with the instantaneous Coulomb
interaction. We emphasize that the fermionic charged fields
$\Psi$ as well as the Lagrange multiplier $A_0$ are {\em gauge
invariant} fields (see Refs.~\cite{boyanrgk,gaugepot}).

The fully renormalized equations of motion for the mean fields can be
obtained by following the
formulation presented in Refs.~\cite{plasmino,boyanfermion}. However, the
counterterms that eliminate the zero temperature
ultraviolet divergences are temperature independent~\cite{lebellac} and
play no r\^ole in the present context, thus we will neglect the zero
temperature renormalization counterterms in our study.

Furthermore we consider a neutral plasma (i.e, with zero chemical
potential for the charged fields) at a temperature
$T\gg m$ with $m$ the (renormalized) mass of the fermions, hence in
what follows we neglect the fermion mass unless otherwise stated.

\subsection{Nonequilibrium Field Theory}

The formulation of nonequilibrium quantum field theory in terms of
the Schwinger-Keldysh or the closed-time-path (CTP) is
standard~\cite{noneq,disip,tadpole}. A path
integral representation requires a contour in the complex time plane,
running forward and
then backwards in time corresponding to the unitary time evolution of
an initially prepared density matrix,
with an effective Lagrangian in terms of fields on the different
branches of the path
$$
{\cal L}_{\rm noneq}
={\cal L}[\Psi^+,\bar\Psi^+,A^+_T,A^+_0]-
{\cal L}[\Psi^-,\bar\Psi^-,A^-_T,A^-_0] \; ,
$$
where the ``$+$'' (``$-$'') superscripts for the fields refer
to fields defined in the forward (backward) time branches.
This description leads to a straightforward diagrammatic expansion of
the nonequilibrium
Green's functions in terms of real-time propagators but with modified Feynman
rules (as compared to standard field theory)~\cite{noneq,disip,tadpole}.
The free fermion and photon propagators are given in detail
in the Appendix.

Since we study two different type of situations: (i) the relaxation of
expectation values in the linearized approximation (linear response)
and (ii) the kinetic equation that describes the evolution of
a non-equilibrium distribution function, we make a distinction between
the real-time propagators used in each case.

\begin{itemize}

\item[(i)]{In the case of relaxation of expectation values in linear
response, the distribution functions that enter
in the fermion and photon propagators (\ref{fermionprop1})-(\ref{gaugeprop2})
are those with {\em equilibrium} distributions.}

\item[(ii)]{When we study the kinetic equations for the expectation
value of the fermion number and photon number operators,
we assume that the initial density matrix is diagonal in the basis of
the quasiparticles (see below) but with nonequilibrium initial distributions.
Thus the propagators obtained with this  initial density matrix have
the free-field form, as given in (\ref{fermionprop1})-(\ref{gaugeprop2})
but the initial occupation numbers are {\em not} the equilibrium
Fermi-Dirac or Bose-Einstein distribution.}
\end{itemize}

\subsection{Real-time relaxation of mean fields in linear response}

The mean fields under consideration are the expectation value of
either fermion or gauge field operators in
the nonequilibrium state induced by the external source. Obviously in
equilibrium and in the absence of external sources these must vanish,
and we introduce time-dependent sources to induce an expectation value
of these operators. Our strategy to study the
relaxation of these mean fields as an initial value problem is to
prepare these mean fields via the adiabatic
switching-on of an external
current\cite{plasmino,boyanfermion,baacke}. Once the current is
switched off the expectation values must relax towards equilibrium and
we study this real-time evolution. This formulation to study the
real-time evolution of mean fields has the advantage that it leads to a
straightforward and systematic implementation
of the dynamical renormalization group method as explained in detail
in Refs.~\cite{boyanrgir,boyanrgk}.

The initial value problem formulation for studying the
linear relaxation of the mean fields   begins by introducing c-number external
sources coupled to the quantum fields.
Let $ \eta(x) $ be the Grassmann-valued fermionic source and ${\vec J}_T(x)$
be the electromagnetic sources,\footnote{In this article we will not discuss
relaxation of the longitudinal gauge field $A_0(x)$ associated with the
instantaneous Coulomb interaction, hence the corresponding external
source is neglected.} then the Lagrangian density becomes
$$
{\cal L}\rightarrow {\cal L}+\bar{\Psi}\eta+\bar{\eta}\Psi
+ {\vec J}_T\cdot{\vec A}_T \;.
$$
The presence of external sources will induce responses of the system.
Let $\Phi(x)$ be a generic quantum field (i.e., fermionic or bosonic)
with $\Phi^{\pm}(x)$ the fields defined on the forward and backward
branches, respectively, and $J(x)$ the corresponding c-number external
source (the same for both branches).
The expectation value of $\Phi(x)$ induced by $ J(x) $ in a linear
response analysis is given by
\begin{eqnarray}
\phi({\vec x},t)&\equiv& \langle\Phi^\pm({\vec
x},t)\rangle_J\nonumber\\ &=&\int d^3x'\int_{-\infty}^{+\infty}
dt' \; G_{\rm ret}({\vec x},t;{\vec x'},t') \; J({\vec x'},t')
\;,\nonumber
\end{eqnarray}
with the retarded Green's function
\begin{eqnarray}
G_{\rm ret}({\vec x},t;{\vec x'},t')&\equiv& i\Big[\left\langle
\Phi^+({\vec x},t)\bar{\Phi}^+({\vec x'},t')\right\rangle
-\left\langle\Phi^+({\vec x},t)\bar{\Phi}^-({\vec
x'},t')\right\rangle\Big]\nonumber\\
&=&i\left\langle\left[\Phi({\vec x},t),\bar{\Phi}({\vec
x'},t')\right]_\mp\right\rangle \theta(t-t') \;,\nonumber
\end{eqnarray}
where expectation values are computed in the CTP formulation in the
full interacting
theory but with vanishing external source, $\theta(t-t')$ is the step
function and
the subscripts $ \mp $ refer to the commutator ($-$) for bosonic or
anticommutator ($+$) for fermionic fields.

A practically useful initial value problem formulation for the
real-time relaxation of mean fields
is obtained by considering that the external source
is adiabatically switched on in time at $t= -\infty$ and suddenly
switched off at $t=t_0$, i.e.,
\begin{equation}
J({\vec x},t)=J({\vec x})\;e^{\epsilon(t-t_0)}\;\theta(t_0-t)\;,
\quad\epsilon\rightarrow 0^+\;.
\label{extsource}
\end{equation}
The adiabatic switching on of the external source induces a mean field
that is dressed adiabatically by the interaction. Then for
$t>t_0$, after the external current has been switched off,
the mean field will relax towards equilibrium, and our aim is
to study this relaxation directly in real time~\cite{baacke}.

The equations of motion of the mean fields are obtained via the tadpole
method~\cite{tadpole} and are automatically causal and retarded.
The central idea of the tadpole method is to write the  field into a
c-numbered expectation  value plus fluctuations
around it, i.e., writing
$$
\Phi^\pm({\vec x},t)=\phi({\vec x},t)+\chi^\pm({\vec x},t) \quad
\mbox{with} \quad \langle\Phi^\pm({\vec x},t)\rangle_J=\phi({\vec
x},t) \; .
$$
The equation of motion for $\phi({\vec x},t)$ is obtained by requiring
$\langle\chi^\pm({\vec x},t)\rangle=0$ to all orders in perturbation theory.

For the study of relaxation of the mean fields, we assume that at
$t=-\infty$ the system is in thermal equilibrium
at a temperature $T$ and henceforth choose $t_0=0$ for convenience.
The retarded and the equilibrium (i.e., time translational invariant)
nature of $G_{\rm ret}({\vec x},t;{\vec x'},t')$
and the adiabatic switching on of $J({\vec x},t)$ entail that
\begin{equation}
\begin{array}{l}
\phi({\vec x},t=0)=\phi_0({\vec x}),\vspace{5pt}\\
\dot\phi({\vec x},t<0)=0 \;,
\end{array}\label{icond}
\end{equation}
where $\phi_0({\vec x})$ is determined by $J({\vec x})$ (or vice versa).
It is worth pointing out that $\dot\phi({\vec x},t)$ is not
specified at $t=0$ although $\dot\phi({\vec x},t<0)=0$.
This is because the external source is
switched off at $t=0$.\footnote{There could be initial time
singularities associated with our choice of the external source
(see Ref.~\cite{baacke}). However the long time asymptotics
is not sensitive to the initial time
singularities~\cite{baacke}, and we will not address this issue here
as it is not relevant for the asymptotic dynamics.}

\subsection{Main ingredients for quantum kinetics}

In Ref.~\cite{boyanrgk} a systematic treatment to obtain the
kinetic equation that determines the real-time evolution of the
distribution function (defined as the expectation value of the number operator)
was established, leading to its interpretation as
a dynamical renormalization group equation.  Here we summarize
the basic steps to obtain the relevant quantum  kinetic equations from
first principles and we refer the reader to
Ref.~\cite{boyanrgk} for more details.

\begin{itemize}

\item[(i)]{Identify the proper degrees of freedom (quasiparticles) to be
described by the kinetic equation, the corresponding microscopic time scale
$\tau_{\rm micro}$ associated with their oscillation, and the number
operator $N({\vec k},t)$ that counts these degrees
of freedom with momentum ${\vec k}$ at time $t$.}

\item[(ii)]{Use the Heisenberg equations of motion to derive a
general rate equation for $d{n}_{\vec k}(t)/dt$
with $n_{\vec k}(t)=\langle N({\vec k},t)\rangle$,
where the expectation value is taken in the initial density matrix.
Assuming that the initial density matrix is diagonal in the basis of
free quasiparticles but with nonequilibrium distribution functions,
we can expand the rate equation in perturbation theory using
nonequilibrium Feynman rules
and free real-time propagators in terms of {\rm nonequilibrium}
distribution functions.
Since the resulting expression is a functional of the distribution
function at initial time,
the solution of the kinetic equation can be obtained by direct integration.

This solution is typically characterized by the emergence of secular terms,
i.e, terms that grow in time in the intermediate asymptotic regime
($\tau_{\rm micro}\ll t\ll \tau_{\rm rel}$) where the scale $\tau_{\rm
rel}$ is that at which perturbation theory breaks
down~\cite{boyanrgir,boyanrgk}.}

\item[(iii)]{These secular terms growing in time become
divergent if the perturbative solution is extrapolated to long times.
However,  perturbation theory is valid and dominated by
the secular terms during the intermediate asymptotic time scales
$\tau_{\rm micro}\ll t\ll \tau_{\rm rel}$. The
dynamical renormalization group is used to resum the secular terms through a
renormalization of the initial distribution functions. This introduces an
arbitrary time scale $\tau$ which serves as a renormalization point.
The $\tau$-independence of the renormalized solution leads to the dynamical
renormalization equation~\cite{boyanrgir,boyanrgk}. This dynamical
renormalization equation describing the evolution of the quasiparticle
distribution is recognized as the quantum kinetic equation.}
\end{itemize}

\section{Photons out of equilibrium}

In this section we study non-equilibrium aspects of photon relaxation
and production. We begin by analyzing the relaxation in real time of a
photon condensate in linear response both in the case of soft $k\lesssim eT$
and semihard (or semisoft) $ eT\ll k \ll T $ assuming the
electromagnetic coupling to be small. We then continue with a study of
the production of semihard photons via off-shell effects.

\subsection{Relaxation of the gauge mean field}

We begin with the relaxation of soft photons of momenta $k\sim eT$.
As mentioned in the previous section, the equation of motion for the
transverse photon mean field can be derived from the tadpole
method~\cite{tadpole} by decomposing
the full quantum fields into c-number expectation values and
quantum fluctuations around the expectation values:
\begin{eqnarray}
{\vec A}^\pm_T({\vec x},t)&=&{\vec a}_T({\vec x},t)+
{\bbox{\cal A}}^\pm_T({\vec x},t),\nonumber\\
{\vec a}_T({\vec x},t)&=&\langle{\vec A}^\pm_T({\vec x},t)\rangle \; .\nonumber
\end{eqnarray}
With the external source $ {\vec J}_T({\vec x},t) $ of the adiabatic
form (\ref{extsource}) chosen to vanish at $ t\geq 0 $,
we find the linear equation of motion for the spatial Fourier
transform of $ {\vec a}_T({\vec x},t) $ for $ t\ge 0 $ to be given by
\begin{equation}
\left(\frac{\partial^2}{\partial t^2}+k^2 \right) \; {\vec a}_T({\vec k},t) +
\int^{t}_{-\infty}dt'\;\Pi_T({\vec k},t-t') \;
{\vec a}_T({\vec k},t')=0 \; ,\label{maxwelleq1}
\end{equation}
where
$$
{\vec a}_T({\vec k},t) \equiv \int d^3x\;e^{-i{\vec k}\cdot{\vec x}} \;
{\vec a}_T({\vec x},t) \; .
$$
Here $\Pi_T({\vec k},t-t')$ is the transverse part of the retarded
photon self-energy and we have made explicit use of its retarded
nature in writing Eq.~(\ref{maxwelleq1}),
which evidently is causal and retarded.

Using the nonequilibrium Feynman rules and the real-time free fermion
propagators given in the Appendix,
we find the transverse polarization to be given by
\begin{eqnarray}
\Pi_T({\vec k},t-t')&=&\int_{-\infty}^{+\infty}
d\omega\;\rho_T(\omega,{\vec k})\sin[\omega(t-t')] \; ,\label{Pit}
\end{eqnarray}
in terms of the spectral density to one-loop order
\begin{eqnarray}
\rho_T(\omega,{\vec k})&=&-e^2\int\frac{d^3 q}{(2\pi)^3}
\big\{[\delta(\omega-p-q)-\delta(\omega+p+q)]
[1+({\vec\hat k}\cdot{\vec\hat p})({\vec\hat k}\cdot{\vec\hat q})]
[1-n_F(p)-n_F(q)]\nonumber\\
&&+\;[\delta(\omega-p+q)-\delta(\omega+p-q)]
[n_F(q)-n_F(p)][1-({\vec\hat k}\cdot{\vec\hat p})
({\vec\hat k}\cdot{\vec\hat q})]\big\} \;,\label{rhot}
\end{eqnarray}
where ${\vec p}={\vec k}+{\vec q}$.

It is now convenient to introduce an auxiliary quantity
$\pi_T({\vec k},t-t')$ as follows
$$
\Pi_T({\vec k},t-t')=\frac{\partial}{\partial t'}\pi_T({\vec k},t-t') \;.
$$
and hence,
$$
\pi_T({\vec k},t-t') = \int_{-\infty}^{+\infty}
 {d\omega\over \omega} \rho_T(\omega,{\vec k})\cos[\omega(t-t')] \;.
$$
Upon integration by parts the equation of
motion takes a form that displays clearly the nature of the initial
value problem:
\begin{eqnarray}
\left(\frac{\partial^2}{\partial t^2}+k^2 \right){\vec a}_T({\vec k},t) &-&
\int^{t}_{0}dt'\;\pi_T({\vec k},t-t')\;\dot{\vec a}_T({\vec k},t')
+\pi_T({\vec k},0)\; {\vec a}_T({\vec k},t)=0\; ,\label{maxwelleq2}
\end{eqnarray}
where a dot denotes derivative with respect to time and we used that $
\dot{\vec a}_T({\vec k},t) = 0 $ for $ t<0 $.
Eq.~(\ref{maxwelleq2}) together with the initial conditions specified at $t=0$
yields a well-defined initial value problem. As discussed above,
the initial value ${\vec a}_T({\vec k},t=0)={\vec a}_T({\vec k},0)$ is
determined by ${\vec J}_T({\vec k},t)$, and we choose
\begin{equation}
\label{condini}
\dot{\vec a}_T({\vec k},t=0)=0
\end{equation}
consistently with the adiabatic switching-on condition (\ref{icond}).

Eq.~(\ref{maxwelleq2}) can be solved by the Laplace transform  method.
We define,
\begin{eqnarray}
\tilde{\vec a}_T(s,{\vec k})&=&\int_0^{\infty} dt\;
e^{-st}\;  {\vec a}_T({\vec k},t)\;,\nonumber\\
\tilde{\pi}_T(s,{\vec k})&=&\int_0^{\infty} dt\;
e^{-st}\; \pi_T({\vec k},t)\;, \label{laplatrans}
\end{eqnarray}
with Re $ \! s>0$. In terms of which the Laplace transform of the
equation of motion leads to
\begin{equation}
\left[s^2+k^2+\tilde{\Pi}_T(s,{\vec k})\right]\tilde{\vec a}_T(s,{\vec k})=
\left[s-\tilde{\pi}_T({s,\vec k})\right]{\vec
a}_T({\vec k},0) \; , \label{LTmaxwelleq1}
\end{equation}
where $\tilde{\Pi}_T(s,{\vec k})$ is the Laplace transform of
${\Pi}_T({\vec k},t-t')$:
$$
\tilde{\Pi}_T(s,{\vec k})=\pi_T({\vec k},0)-s\; \tilde{\pi}_T(s,{\vec k})\;.
$$
The solution of Eq.~(\ref{LTmaxwelleq1}) is given by
\begin{equation}
\tilde{\vec a}_T(s,{\vec k})=\frac{1}{s}\left\{1-
\tilde{\Delta}_T(s,{\vec k})[k^2+\pi_T({\vec k},0)]\right\}
{\vec a}_T({\vec k},0) \label{laplasol}\;,
\end{equation}
where the retarded transverse photon propagator
$\tilde{\Delta}_T(s,{\vec k})$ is given by
\begin{equation}
\tilde{\Delta}_T(s,{\vec k})=[s^2+k^2+
\tilde{\Pi}_T(s,{\vec k})]^{-1}\label{deltatil} \;.
\end{equation}
The real-time evolution of ${\vec a}_T({\vec k},t)$ is obtained by performing
the inverse Laplace transform along the Bromwich contour which is to
the right of all singularities of $\tilde{\vec a}_T(s,{\vec k})$
in the complex $s$-plane.

We note that this result is rather {\em different} from that obtained
in Ref.~\cite{boyanhtl} in the structure of the solution,
in particular the prefactor $1/s$ in Eq.~(\ref{laplasol}).
This can be traced back to the
{\em different} manner in which we have set up the initial value
problem, as compared to the case studied in~\cite{boyanhtl}.
The adiabatic source (\ref{extsource}) (for $t_0=0$)
has a Fourier transform that has a simple pole in the frequency plane,
this translates into the $1/s$ factor in Eq.~(\ref{laplasol}).
This particular case was {\em not} contemplated in those studied in
Ref.~\cite{boyanhtl} since the source (\ref{extsource}) is
{\em not} a regular function of the frequency.
This difference will be seen to be at the heart of important
aspects of relaxation of the mean field condensate in the soft
momentum limit as discussed in detail below.

It is straightforward to see that the residue vanishes
at $ s=0 $ in the solution (\ref{laplasol}). The singularities of
$ \tilde{\vec a}_T(s,{\vec k}) $ are those arising from the retarded
transverse photon propagator $\tilde{\Delta}_T(s,{\vec k})$.

\subsubsection{Soft photons $k \sim eT$: real-time Landau damping}

For soft photons of momenta $k\sim eT$, the leading ($\sim e^2T^2$)
contribution to the photon self-energy arises from loop momenta of
order $T$ and is referred to as a hard thermal loop
(HTL)~\cite{htl,rob2}.  In this
region of momenta, the contribution of the fermion loop is non-perturbative.

In the HTL approximation ($s\sim k\ll q$), after some algebra
we find that the HTL contribution to $\tilde\Pi_T(s,{\vec k})$
arises exclusively from the terms associated with $\delta(\omega\mp p\pm q)$
in Eq.~(\ref{rhot}) and reads
\begin{equation}\label{HTLPis}
\tilde\Pi_T(s,{\vec k})=
\frac{e^2 T^2}{12}\left[\frac{is}{k}
\left(1+\frac{s^2}{k^2}\right)\ln\frac{is+k}{is-k}-
2 \; \frac{s^2}{k^2}\right] \;.
\end{equation}
The delta functions $\delta(\omega\mp p\pm q)$ have support below the
light cone
(i.e., $\omega^2 < k^2$) and correspond to Landau damping
processes~\cite{rob3} in which the soft photon scatters a
hard fermion in the plasma.

The analytic continuation of $\tilde\Pi_T(s,{\vec k})$ in the complex
$s$-plane is defined by
\begin{eqnarray}
\Pi_T(\omega,{\vec k})&\equiv&\tilde{\Pi}_T(s=-i\omega\pm 0^+,{\vec k})
=\mbox{Re}\Pi_T(\omega,{\vec k})\pm i\;\mbox{Im}\Pi_T(\omega,{\vec k}).
\label{pi:anacont}
\end{eqnarray}
From Eq.~(\ref{Pit}) it is straightforward to see that the real and
imaginary parts are related by the usual dispersion relation.
The analytical continuation of $\tilde\Delta_T(s,{\vec k})$ and
$\tilde\pi_T(s,{\vec k})$
can be defined analogously, and they are related to
$\Pi_T(\omega,{\vec k})$ by
\begin{eqnarray*}
\Delta_T(\omega,{\vec k})&=&-[\omega^2-k^2-\Pi_T(\omega,{\vec k})]^{-1}\;,\\
\mbox{Re}\Pi_T(\omega,{\vec k})&=&\pi_T({\vec k},0)-
\omega\;\mbox{Im}\pi_T(\omega,{\vec k})\;,\\
\mbox{Im}\Pi_T(\omega,{\vec k})&=&\omega\;\mbox{Re}\pi_T(\omega,{\vec k})\;.
\end{eqnarray*}
The analytical continuation of $\tilde\Pi_T(s,{\vec k})$ in the HTL
limit thus reads
\begin{eqnarray}
\Pi_T(\omega,{\vec k})&=&\frac{e^2
T^2}{12}\left[2\frac{\omega^2}{k^2}+\frac{\omega}{k}
\left(1-\frac{\omega^2}{k^2}\right)
\ln\left|\frac{\omega+k}{\omega-k}\right|\right]\nonumber\\
&&-\;i\frac{\pi e^2
T^2}{12}\frac{\omega}{k}\left(1-\frac{\omega^2}{k^2}\right)
\theta(k^2-\omega^2) \;. \nonumber
\end{eqnarray}
In the HTL limit it is straightforward to prove that $\pi_T({\vec k},0)=0$.
We note that $\mbox{Im}\Pi_T(\omega,{\vec k})$ only has support below
the light cone and vanishes {\em linearly} as $\omega\rightarrow k$ from below.
Since for soft momenta $k\ll eT$ the HTL photon self-energy is
comparable in magnitude to or larger than the free inverse propagator,
it has to be treated nonperturbatively.

Isolated poles of $\Delta_T(\omega,{\vec k})$ in the complex $\omega$-plane
correspond to collective excitations. The transverse photon pole
$\omega_T(k)$ is real and determined by~\cite{lebellac}
$$
\omega^2_T(k)-k^2-\mbox{Re}\Pi_T[\omega_T(k),{\vec k}]=0\;.
$$
For ultrasoft photons $k\ll eT$, the dispersion relation of the
collective excitations is~\cite{lebellac}
$$
\omega_T^2(k)= \omega_{P}^2 + \frac{6}{5} k^2+{\cal O}\left( {  k^4 \over
e^2 \, T^2 }\right) \; ,
$$
where $\omega_{P}=eT/3$ is the plasma frequency.
Consequently, the retarded transverse photon propagator
$\tilde{\Delta}_T(s,{\vec k})$
has two isolated poles at $s=\pm i\omega_T(k)$ and a branch cut from
$s=-ik$ to $s=ik$.

Having analyzed the analytic structure of $\tilde{\Delta}_T(s,{\vec k})$,
we can now perform the integral along the Bromwich contour to obtain the
real-time evolution of
${\vec a}_T({\vec k},t)$. Closing the contour in the half-plane
$\mbox{Re}\;\!s<0$, we find ${\vec a}_T({\vec k},t)$ for $t>0$ to be given by
$$
{\vec a}_T({\vec k},t)={\vec a}^{\it pole}_T({\vec k},t)+
{\vec a}^{\it cut}_T({\vec k},t) \;,
$$
with
\begin{eqnarray}
{\vec a}^{\it pole}_T({\vec k},t)&=& \frac{k^2Z_T(k)}{\omega_T^2(k)}
\cos[\omega_T(k)t]\;{\vec a}_T({\vec k},0) \;,
\label{sol:aTpole}\\
{\vec a}^{\it cut}_T({\vec k},t)&=& k^2\int_{-k}^{+k}\frac{d\omega}{\omega}\;
\beta_T(\omega,k)\;e^{-i\omega t}\;{\vec a}_T({\vec k},0),\label{sol:aTcut}
\end{eqnarray}
where
\begin{eqnarray}
Z_T(k)&=&\left[1-\frac{\partial\mbox{Re}\Pi_T(\omega,{\vec
k})}{\partial\omega^2} \right]^{-1}_{\omega=\omega_T(k)}\; ,\cr \cr
\beta_T(\omega,k)&=&\frac{\frac{e^2 T^2}{12}\frac{\omega}{k}
\left(1-\frac{\omega^2}{k^2}\right)}
{\left\{\omega^2-k^2-\frac{e^2 T^2}{12}
\left[\frac{2\omega^2}{k^2}+\frac{\omega}{k}
\left(1-\frac{\omega^2}{k^2}\right)
\ln\frac{k+\omega}{k-\omega}\right]\right\}^2+
\left[\frac{\pi e^2 T^2}{12}\frac{\omega}{k}
\left(1-\frac{\omega^2}{k^2}\right)\right]^2} \;. \label{gorda}
\end{eqnarray}
The solution evaluated at $t=0$ must match the initial condition, this
requirement leads immediately to
the  sum rule~\cite{lebellac}
$$
\int_{-\infty}^{+\infty}
\frac{dq_0}{q_0}\;\tilde{\rho}_T(q_0,q)=\frac{1}{q^2},
$$
where $ \tilde{\rho}_{T}(q_0,q) $ is the HTL-resummed spectral density
for the transverse photon
propagator:\footnote{In our notation the spectral density for the self-energy
is denoted by $\rho$, and the spectral density for the corresponding
propagator is denoted by $\tilde\rho$.}
\begin{eqnarray}
\tilde{\rho}_{T}(q_0,q)&=&\frac{1}{\pi}\mbox{Im}\Delta_T(q_0,q)\nonumber\\
&=&\mbox{sgn}(q_0)\,Z_{T}(q)\,\delta[q^2_0-\omega^2_{T}(q)]+
\;\beta_{T}(q_0,q)\; \theta(q^2-q_0^2) \;.
\label{tilderhot}
\end{eqnarray}
A noteworthy feature of  the cut contribution (\ref{sol:aTcut}) is the
factor $\omega$ in the denominator. The presence
of this factor can be traced back to the preparation of the initial
state via the adiabatically switched-on external source,
which is the switched off at $t=0$. The Fourier transform of this
source is proportional to $1/\omega$
and results in the prefactor of $\beta_T(\omega,k)$ in Eq.~(\ref{sol:aTcut}).

For $k\sim eT$, Eq.~(\ref{sol:aTcut}) cannot be evaluated in
closed form but its long time asymptotics can be extracted by writing the
integral along the cut as a contour integral in the complex $\omega$-plane.
The integration contour ${\cal C}$ is chosen to run clockwise along the
segment $-k<\omega< k$ on the real axis, the line $\omega=k-iz$ with
$0\le z<\infty$, then around an arc at infinity and back to the real axis
along the line $\omega=-k-iz$. After some algebra we obtain the
following expression
\begin{eqnarray*}
{\vec a}^{\it cut}_T({\vec k},t)
&=&k^2\Bigg\{e^{-ikt}\int_0^\infty dz\,e^{-zt}\,
\frac{i\beta_T(k-iz,k)}{k-iz}+\mbox{ c.c.}\\
&&-\;2\pi i\!
\sum_{\hbox{\scriptsize poles}\atop\hbox{\scriptsize inside ${\cal C}$}}
\text{Res}\!\left[\frac{\beta_T(\omega,k)}{\omega}e^{-i\omega t}\right]\Bigg\}
 \; {\vec a}_T({\vec k},0) \;.
\end{eqnarray*}
Because of the exponential factor $e^{-zt}$ in the integrals ,
the dominant contributions to the integral at long times ($t\gg 1/k$)
arise from the end points of the branch cut with $z\ll k$, and lead to
a long time behavior characterized by a power law:
\begin{equation}
{\vec a}^{cut}_T({\vec k},t)  \buildrel{kt\gg 1}\over=
-\frac{12}{e^2 T^2}\frac{\cos kt}{t^2}\;{\vec a}_T({\vec k},0)\left[1
+ {\cal O}\left({1 \over t} \right)\right] \;,   \label{tail}
\end{equation}
which confirms the results of Ref.~\cite{boyanhtl}. The contribution from
the poles inside the contour ${\cal C}$ is dominated at long times by
the pole closest to the real axis.

\begin{center}
\begin{figure}
\epsfig{file=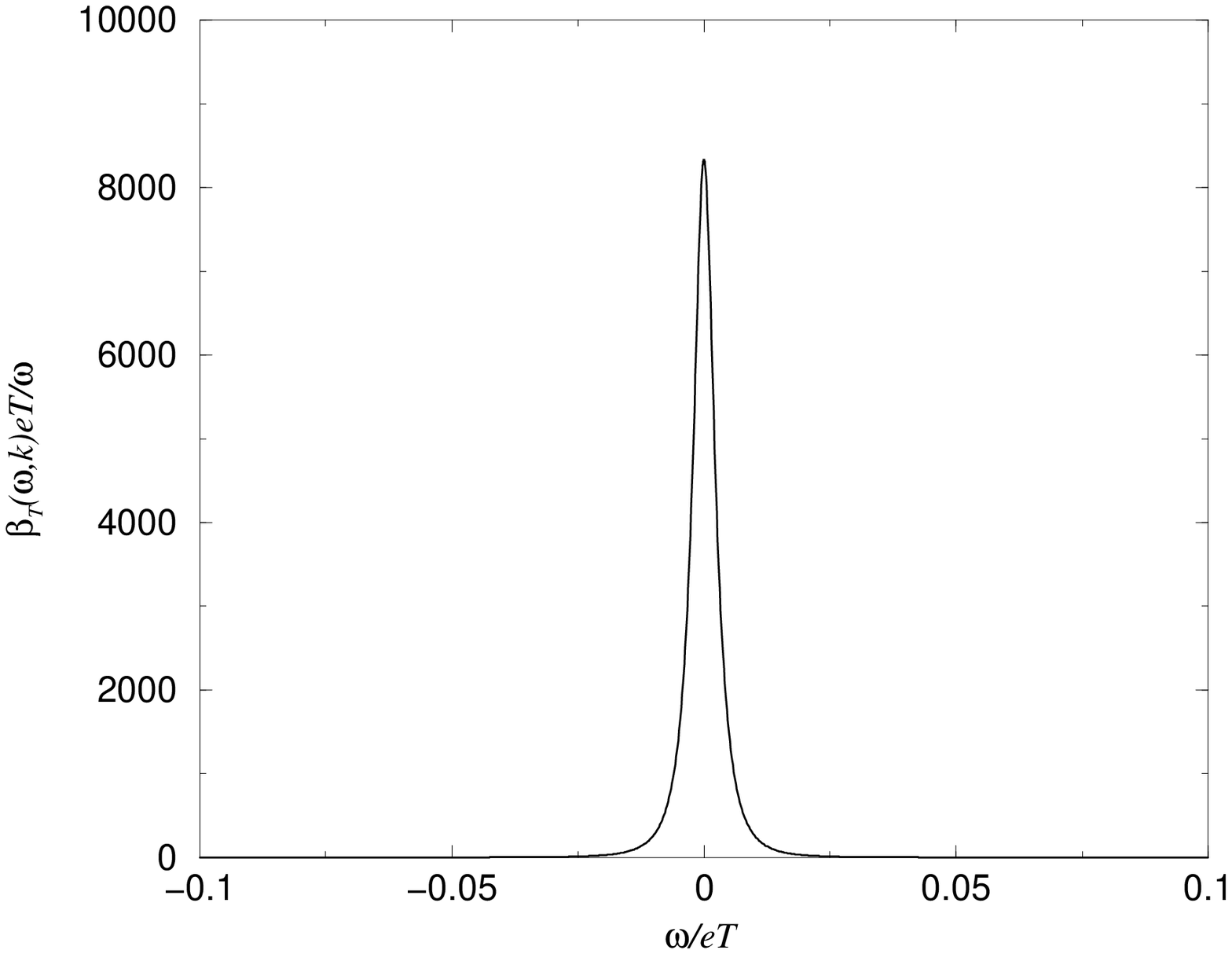,width=12.0cm,height=8.0cm}
\vspace{.1in}
\caption{$\beta_T(\omega,k)/\omega$ vs $\omega$ for ultrasoft photon
of momentum  $k/eT=0.1$.
\label{fig:softphoton}}
\end{figure}
\end{center}

For very soft momenta $ k\ll eT $ the function
$\beta_T(\omega,k)/\omega$ is strongly peaked at $\omega=0$ as
depicted in Fig.~\ref{fig:softphoton}. For $ k\ll eT$ and $k \gg \omega$,
Eq.~(\ref{gorda}) takes a Breit-Wigner form
\begin{equation}
\frac{1}{\omega}\beta_T(\omega,k)\buildrel{\omega\ll k\ll eT}\over=
\frac{1}{\pi k^2}\frac{\Gamma_k}{\omega^2+\Gamma^2_k}
\quad\mbox{with}\quad\Gamma_k=\frac{12k^3}{\pi e^2 T^2} \;. \label{breitwigner}
\end{equation}
Using this narrow-width approximation to evaluate Eq.~(\ref{sol:aTcut})
yields,
\begin{equation}
{\vec a}^{\it cut}_T({\vec k},t) \buildrel{kt\gg 1}\over=
\left[ e^{-\Gamma_k t} - {2 \,
\Gamma_k \over \pi \, k^2 \; t} \sin kt + {\cal O}\left(\Gamma_k \over
k^3 \; t^2 \right) \right]\;{\vec a}_T({\vec k},0) \;.
\label{landaudamping}
\end{equation}
The end-point contribution which results in a power law as in Eq.(\ref{tail})
is very small compared with Eq.~(\ref{landaudamping}) for time
$ t\lesssim 1/\Gamma_k $.
This power law becomes dominant at time scales {\em much longer} than
$1/\Gamma_k$, because its amplitude at $t\sim 1/\Gamma_k$ is of order
$(k/eT)^6\ll 1$ in the soft momentum limit.

Thus, for adiabatically prepared mean fields of soft momentum $k \ll
eT$ we find the real-time behavior
$$
{\vec a}_T({\vec k},t) \buildrel{kt \gg 1\, , \, \Gamma_k \, t
\lesssim 1}\over=
{\vec a}_T({\vec k},0)\left\{\frac{k^2Z_T(k)}{\omega_T^2(k)}
\cos[\omega_T(k)t]+ e^{-\Gamma_k t}\right\} \;.
$$

It is worth emphasizing that the exponential decay is not the same as
the usual decay of an unstable particle
(or even collisional broadening) for which the amplitude relaxes to
zero at times longer than the relaxation time. In this
case the asymptotic behavior of the amplitude is completely determined
by the transverse photon pole $\omega_T(k)$, and to this order in the
HTL approximation the collective excitation is stable.
The exponential relaxation does not arise from a resonance at the
position of the transverse photon pole but at zero frequency. Clearly
we expect a true exponential damping of the collective excitation at
higher order as a result of collisional broadening.

Our results thus confirm those found in hot scalar quantum
electrodynamics~\cite{rajantie}
wherein a numerical analysis of the relaxation
of the mean field was carried out. In that reference the authors
studied the real time evolution in terms of local
Hamiltonian equations obtained by introducing a non-local
auxiliary field. The initial conditions chosen there for
this non-local auxiliary field correspond precisely to the choice of
an adiabatically switched on current leading to an
adiabatically prepared initial value problem.

We emphasize that the exponential relaxation (\ref{landaudamping}) can
{\em only} be probed by adiabatically switching on an external source
[see Eq.~(\ref{extsource})]
whose temporal Fourier transform has a simple pole at $\omega=0$ which is the
origin of the prefactor $1/\omega$ in Eq.~(\ref{sol:aTcut}). That is the
adiabatic preparation of the initial state excites this pole in the
Landau damping region. For external sources whose temporal Fourier transform
is regular at $\omega=0$ no exponential relaxation
arises, in agreement with the results of Ref.~\cite{boyanhtl}.
Thus we find that the exponential relaxation is {\em not}
a generic feature of the evolution of the mean field, but emerges only
for particular (albeit physically motivated) initial conditions.

\subsubsection{Semi-hard photons $eT\ll k \ll T$: anomalous relaxation}

Most of the studies of the photon self-energy in the hard thermal loop
approximation focused on the soft external momentum region $k \ll eT$
(with $e\ll 1$)~\cite{htl,rob2,rob3,weldon1,robinfra,lebellac} wherein
the contribution of the hard thermal fermion loop must be treated
non-nonperturbatively.

However there are important reasons that warrant a consideration of
the region of {\em semihard} photons $eT\ll k\ll T$.
From a phenomenological standpoint, hard and semihard
photons produced in the QGP phase are
important electromagnetic probes of the deconfined
phase~\cite{qgp1,qgp2,kapusta,baier} hence the study of all aspects of
the space-time evolution, production and relaxation of photons  to
explore potential experimental signatures is warranted.

A more theoretical justification of the relevance of this region is
that whereas the photon self-energy for this region of momenta is
still dominated by the hard thermal fermion loop contribution
(\ref{HTLPis}), its contribution to the full photon propagator is now
{\em perturbative} as compared with the free inverse propagator. The
validity of perturbation theory in this regime licenses us to study
the real-time evolution with the tools developed in
refs.~\cite{boyanrgir,boyanrgk} that provide a consistent
implementation of the renormalization group to study real-time
phenomena.

\begin{center}
\begin{figure}
\epsfig{file=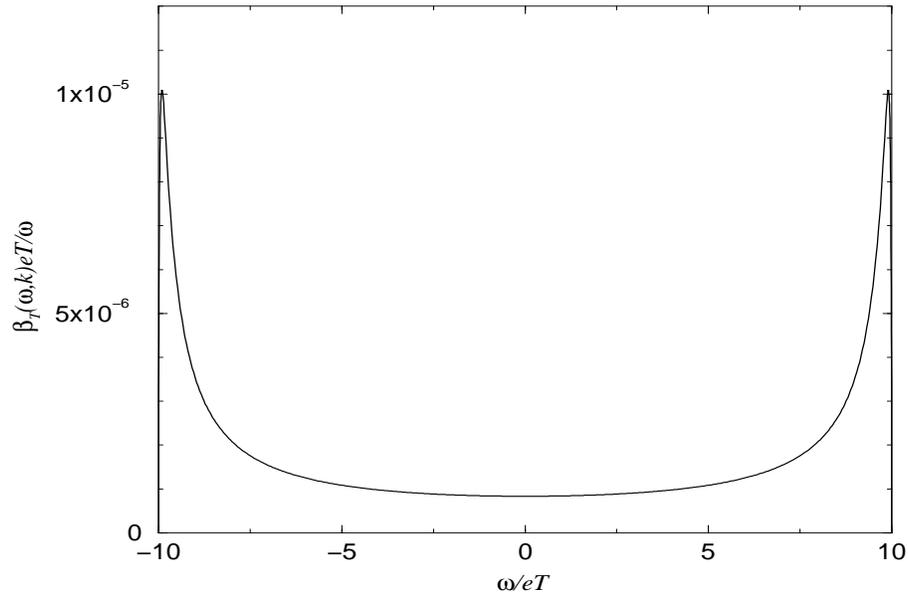,width=12.0cm,height=8.0cm}
\vspace{.1in}
\caption{$\beta_T(\omega,k)/\omega$ vs $\omega$ for semihard photon
of momentum  $eT/k=0.1$.
\label{fig:semihardphoton}}
\end{figure}
\end{center}

The dynamical renormalization group approach introduced in
Refs.~\cite{boyanrgir,boyanrgk} is particularly suited to
study the real-time evolution in the case in which there are
(infrared) threshold singularities in the spectral density.
To understand the potential emergence of anomalous thresholds in the
semihard and hard limit for the photon self-energy
we note that at leading order in HTL the Laplace transform of the
inverse photon propagator Eqs.~(\ref{deltatil})-(\ref{HTLPis}) reads
$$
\tilde\Delta^{-1}(s,{\vec k})=(s^2+k^2)
\left[1+\frac{e^2 T^2}{12}\frac{is}{k^3}\ln\frac{is+k}{is-k}\right]
- \frac{e^2 T^2}{6}\frac{s^2}{k^2} \;.
$$
The transverse photon poles are for $s=\pm i[k+e^2T^2/12k+{\cal
O}(T^4/k^3)]$ which in
the limit $k\gg eT$ approach $s\rightarrow \pm ik$. That is,
\begin{equation}\label{omesemid}
\omega_T(k)= k+\frac{e^2T^2}{12k}+{\cal O}\left(\frac{T^4}{k^3}\right)\;.
\end{equation}
In this limit the poles merge with the thresholds of the
logarithmic branch cut from Landau damping and are no longer isolated
from the continuum.
The approach of the pole to the Landau
damping threshold and the enhancement of the spectral density
near threshold for $k\gg eT$ is clearly displayed in
Fig.~\ref{fig:semihardphoton}.
While the perturbative expansion in the effective coupling $e^2T^2/k^2$ is
warranted in the semihard case under consideration, the wave function
renormalization
constant evaluated as the residue at the pole is a non-analytic
function of this
coupling and given by
$$
Z_T(k) = 1+ \frac{e^2T^2}{12 k^2}\left[ \ln\left(
\frac{e^2T^2}{24k^2}\right)+ 3 +
{\cal O}\left( \frac{e^2 T^2}{k^2}\ln \frac{eT}{k}\right)\right]\; .
$$
This is obviously a consequence of the logarithmic threshold
singularities in the semihard regime.
This threshold singularity is reminiscent of those studied in detail
in Ref.\cite{boyanrgir}.
In this reference it was shown that the Bloch-Nordsieck
resummation which is equivalent to a renormalization group improvement
of the space-time Fourier transform of the self-energy, leads to a
real-time evolution which is obtained by the implementation of the
dynamical renormalization group in real time~\cite{boyanrgir}. Since
the infrared threshold singularities in the semihard limit are akin to
those studied in Ref.~\cite{boyanrgir}, we now implement the
dynamical renormalization group program advocated in that reference to
study the relaxation of the photon mean field in the semihard limit
directly in real time.

Perturbation theory in terms of the effective coupling $e^2T^2/k^2$
is in principle reliable for semihard photons,
hence we can try to solve Eq.~(\ref{maxwelleq2}) by perturbative expansion
in powers of $e^2$ (the true dimensionless coupling is $e^2T^2/k^2$).
Writing
\begin{eqnarray*}
{\vec a}_T({\vec k},t)&=&{\vec a}^{(0)}_T({\vec k},t)+
e^2 \;{\vec a}^{(1)}_T({\vec k},t)+{\cal O}(e^4) \;,\\
\pi_T({\vec k},t-t')&=&e^2 \; \pi_T^{(1)}({\vec k},t-t')+{\cal O}(e^4),
\end{eqnarray*}
and expanding consistently in powers of $e^2$ we obtain a hierarchy of
equations:
\begin{eqnarray*}
\left(\frac{\partial^2}{\partial t^2}+k^2 \right){\vec
a}^{(0)}_T({\vec k},t)&=&0,\\
\left(\frac{\partial^2}{\partial t^2}+k^2 \right){\vec
a}^{(1)}_T({\vec k},t)&=&
\int^{t}_{0}dt'\;\pi^{(1)}_T({\vec k},t-t')\;\dot{\vec
a}^{(0)}_T({\vec k},t') \;,\\
\vdots\quad\quad&&\quad\quad\vdots
\end{eqnarray*}
where we have explicitly used the fact that $\pi_T({\vec k},0)=0$
in the HTL limit.
Starting from the solution to the zeroth-order equation
\begin{equation}\label{solord0}
{\vec a}^{(0)}_T({\vec k},t)=\sum_{\lambda=1}^2\left[A_\lambda({\vec
k})\;e^{-ikt} +A^{\ast}_\lambda({\vec k})\;e^{ikt}\right]{\bbox{\cal
E}}_\lambda({\vec k}) \; ,
\end{equation}
with ${\bbox{\cal E}}_\lambda({\vec k})$ being the polarization vector,
and the retarded Green's function of the unperturbed problem:
$$
G^{(0)}_{\rm ret}(k,t-t')=\frac{\sin[k(t-t')]}{k}\;\theta(t-t') \;,
$$
the solution to the first-order equation reads
\begin{eqnarray}
e^2{\vec a}^{(1)}_T({\vec k},t)&=&-\frac{i}{4}\sum_{\lambda=1}^2
A_\lambda({\vec k}) \; {\bbox{\cal E}}_\lambda({\vec k})
\int_{-\infty}^{+\infty} d\omega\; \frac{\rho_T(\omega,{\vec k})}{\omega}
\; \bigg\{\bigg[\frac{e^{-ikt}}{\omega-k}
\left(t-\frac{1-e^{-i(\omega-k)t}}{i(\omega-k)}\right)
\nonumber\\&&+\;\frac{e^{ikt}}{\omega+k}
\left(\frac{1-e^{-2ikt}}{2ik}+\frac{1-e^{i(\omega-k)t}}{i(\omega-k)}
\right)\bigg]
+\;(\omega\rightarrow -\omega)\bigg\}+\mbox{ c.c.} \;,\label{aTsol1}
\end{eqnarray}
where
\begin{eqnarray*}
\rho_T(\omega,{\vec k})&=&\frac{1}{\pi}\mbox{Im}\Pi_T(\omega,{\vec k})
=\frac{e^2T^2}{12}\frac{\omega}{k}\left(1-\frac{\omega^2}{k^2}\right)
\theta(k^2-\omega^2).
\end{eqnarray*}
Potential secular terms (that grow in time) arise at long times from
the regions in $\omega$ at which the denominators in Eq.~(\ref{aTsol1})
are resonant. These are extracted in the long time limit by using the
formulae in an appendix of Ref.~\cite{boyanrgir}. Particularly
relevant to the case under consideration are the following~\cite{boyanrgir}:
\begin{eqnarray}\label{formulas1}
\int_0^{\infty} {dy \over y^2} \left(1 -\cos yt\right) p(y)
&\buildrel{t \to\infty}\over=&
\frac{\pi}2 \; t \; p(0) + p'(0) \;\left[\ln(\mu\,t)+\gamma_E
\right]\nonumber\\
&& + \int_0^{\infty} {dy \over
y^2}\left[p(y)-p(0)-y\;p'(0)\;\theta(\mu -y)\right]
+ {\cal O}\left(t^{-1}\right) \;, \nonumber\\
\int_0^{\infty}{ dy \over y}\left(t-{\sin yt \over y }\right)
p(y) &\buildrel{t \to\infty}\over=&
t \; p(0) \left[ \ln(\mu\,t) + \gamma_E - 1 \right]\nonumber\\
&&\;+\;t\;\,{\rm PV}\int_0^{\infty} {dy \over y}
\left[p(y)-p(0)\;\theta(\mu - y)\right]+{\cal O}\left(t^{-1}\right) \;,
\end{eqnarray}
where $ \gamma_E = 0.5772157\ldots $ is the Euler-Mascheroni
constant and as can be easily shown the dependence on $\mu$ in the
above integrals cancels. For our analysis, the integration variable
$y= \omega \pm k$ and $ p(y) = \rho_T(\omega,{\vec k})/\omega$.
The terms that grow linearly in time are recognized as
those emerging in Fermi's golden rule from
elementary time-dependent perturbation theory in quantum
mechanics. We note that $p(0)=0$ and $p'(0)\neq 0$,
thus the first integral above gives a contribution to
the {\em real part} of the mean field (i.e., the amplitude)
that features a {\em logarithmic} secular term, whereas the second
integral contributes to the {\em imaginary part} (i.e.,  the oscillating phase)
with a linear secular term, which as will be recognized below determines a
perturbative shift of the oscillation frequency.

Substituting $\rho_T(\omega,{\vec k})$ into Eq.~(\ref{aTsol1}), we obtain
\begin{eqnarray}
e^2{\vec a}^{(1)}_T({\vec k},t)&=&-\sum_{\lambda=1}^2 A_\lambda({\vec k}) \;
{\bbox{\cal E}}_\lambda({\vec k})e^{-ikt}\;\bigg[\frac{e^2T^2}{12k^2}
(\ln 2kt+\gamma_E-1) + i\,\delta_k\, t+
{\cal O}\left(t^{-1}\right)\bigg]+\mbox{c.c.} \;,\label{aTsol1s}
\end{eqnarray}
where
\begin{equation}
\delta_k \equiv\left.\frac{\mbox{ Re}\Pi_T(\omega,{\vec k})}{2\omega}
\right|_{\omega=k} =\frac{e^2 T^2}{12k} \;.\label{masshift}
\end{equation}
Secular divergences are an ubiquitous feature in the perturbative
solution of differential equations with oscillatory behaviour. A resummation
method that improves the asymptotic solution of these differential
equations was introduced in Ref.~\cite{DRG} which is interpreted as a dynamical renormalization group.
This method is very powerful and not only allows a consistent resummation of the secular terms in the
perturbative expansion but provides also a consistent reduction of the dynamics to the slow degrees of freedom
as is explicitly explained in the work of Chen et. al. and Kunihiro et. al. in\cite{DRG}.
 This method was
recently  generalized to the realm of quantum field theory as a
dynamical renormalization group to resum the perturbative series for the
real time  evolution of non-equilibrium expectation values~\cite{boyanrgir,boyanrgk}. This generalization
of the work of reference\cite{DRG} for differential equations  to  non-equilibrium
quantum field theory is a major step since in non-equilibrium systems the equations of motion for expectation
values are non-local and as shown in detail in this work require a resummation of Feynman diagrams.

While the {\em linear} secular terms have a natural interpretation in terms of
renormalization of the mass (the imaginary part)
or a quasiparticle width (the real part)~\cite{boyanrgir},
the {\em logarithmic} secular term found above is akin to those found in
Refs.~\cite{boyanrgir,boyanrgk} that lead to anomalous relaxation.
Furthermore the origin of these logarithmic
secular terms is similar to the threshold infrared divergences and
threshold enhancement of the spectral density
due to the presence of nearby poles~\cite{boyanrgir,boyanrgk}.

Thus following the method presented in Refs.~\cite{boyanrgir,boyanrgk} we
implement the dynamical renormalization group by introducing a (complex)
amplitude renormalization factor in the following manner,
\begin{eqnarray}
A_\lambda({\vec k})&=&{\cal Z}_k(\tau)\;{\cal A}_\lambda({\vec
k},\tau)\; , \nonumber
\end{eqnarray}
where
${\cal Z}_k(\tau)=1+e^2 z_k^{(1)}(\tau)+{\cal O}(e^4)$ is a
multiplicative renormalization constant, ${\cal A}_\lambda({\vec k},\tau)$
is the renormalized initial value, and $\tau$ is an
arbitrary renormalization scale at which the secular divergences are
cancelled~\cite{boyanrgir,boyanrgk}.
Choosing
$$
e^2 z_k^{(1)}(\tau)=\frac{e^2T^2}{12k^2}(\ln 2k\tau +\gamma_E-1)+
i\;\delta_k\; \tau \; ,
$$
we obtain to lowest order in $e^2T^2/k^2$ that the solution of the
equation of motion is given by
\begin{eqnarray}
{\vec a}_T({\vec k},t)&=&\sum_{\lambda=1}^2 {\cal A}_\lambda({\vec k},\tau)
\;{\bbox{\cal E}}_\lambda({\vec k})\;e^{-ikt}
\;\left[1-\frac{e^2T^2}{12k^2}\ln\frac{t}{\tau}
-i\delta_k(t-\tau)\right]+\mbox{ c.c.}\nonumber\\
&&+\;\text{nonsecular terms} \;,\label{aTren}
\end{eqnarray}
which remains bounded at large times $t$ provided that $\tau$ is chosen
arbitrarily close to $t$.
The solution does not depend on the renormalization scale
$\tau$ and this independence leads to the dynamical renormalization group
equation\cite{boyanrgir,boyanrgk}, which to this order is given by
$$
\left[\frac{\partial}{\partial\tau}+\frac{e^2T^2}{12k^2\tau}+
i\; \delta_k\right]{\cal A}_\lambda({\vec k},\tau)=0 \;,
$$
with the solution
$$
{\cal A}_\lambda({\vec k},\tau)={\cal A}_\lambda({\vec
k},\tau_0)\; e^{-i\delta_k(\tau-\tau_0)}
\left(\frac{\tau}{\tau_0}\right)^{-\frac{e^2T^2}{12k^2}} \;,
$$
where $\tau_0$ is the time scale such that this
intermediate asymptotic solution is valid and physically corresponds
to a microscopic scale, i.e, $ \tau_0\sim 1/k $.
Finally, setting $ \tau=t $ in Eq.~(\ref{aTren}) we find that ${\vec
a}_T({\vec k},t) $ evolves at intermediate asymptotic times $ t \gg 1/k $  as
\begin{equation}
{\vec a}_T({\vec k},t)=\sum_{\lambda=1}^2{\cal A}_\lambda({\vec k},\tau_0)
\;{\bbox{\cal E}}_\lambda({\vec k})\;e^{-i(k+\delta_k)(t-\tau_0)}
\left(\frac{t}{\tau_0}\right)^{-\frac{e^2T^2}{12k^2}}+\rm{c.c.} \;.
\label{finalsol}
\end{equation}
From Eq.~(\ref{masshift}) we see that $\delta_k$ is consistent with the
photon thermal mass $m^2_{\gamma}= e^2 T^2/6$~\cite{lebellac}
for $k^2 \gg m^2_{\gamma}$.

As discussed in detail in Ref.~\cite{boyanrgir} the dynamical
renormalization group solution (\ref{finalsol}) is also
obtained via the Fourier transform of the renormalization group
improved propagator in frequency-momentum space, hence the
above solution corresponds to a renormalization group improved {\em
resummation} of the self-energy~\cite{boyanrgir,boyanrgk}.

This novel anomalous power law relaxation of the photon mean field
will be confirmed below in our study of the
kinetics of the photon distribution function in the linearized approximation.

This anomalous power law relaxation is obviously very slow in the
semihard regime in which the HTL approximation and
perturbation theory is valid.
At higher orders we expect exponential relaxation due to collisional processes,
which emerges from linear secular terms~\cite{boyanrgir,boyanrgk}
in a perturbative solution of the real-time equations of motion.
The power law relaxation will then compete with the exponential
relaxation and we expect a crossover time scale
at which relaxation will change from a power law to an exponential.
Clearly an assessment of this time scale
requires a detailed calculation of higher order contributions which we
expect to study in a forthcoming article.

A related crossover of behavior will be found below for the evolution
of the photon distribution function and photon production.

\subsubsection{Secular terms from the Laplace transform for semihard photons}

We show here how secular terms for semihard photons ($eT \ll k \ll T$)
can be obtained from the Laplace transform representation of the
photon mean field given by Eqs.~(\ref{sol:aTcut})-(\ref{gorda})
for large times $kt\gg 1 $.

The integrand in Eq.~(\ref{sol:aTcut}) can be expanded as follows for
semihard momentum,
$$
\frac{1}{\omega}\beta_T(\omega,k)\buildrel{eT\ll k\ll T}\over=
 -{ e^2 T^2 \; \over 12 k^3} \; {1 \over
\omega^2-k^2} \left[1 + {\cal O}\left({e^2 T^2\over k^2}\right)\right] \;.
$$
Inserting this expansion in Eq.~(\ref{sol:aTcut}) gives for
${\vec a}^{\it cut}_T({\vec k},t) $,
$$
{\vec a}^{\it cut}_T({\vec k},t) = - { e^2 T^2\over 12 k} \;
\mbox{PV} \int_{-k}^{+k}
{d \omega \over \omega^2-k^2} \; e^{-i \omega t} \;
{\vec a}_T({\vec k},0) \;
\left[1 + {\cal O}\left( {e^2 T^2\over k^2}\right) \right] \;.
$$
where PV stands for the principal value prescription.
This integral yields for asymptotic times to first order in
${e^2 T^2/k^2}$
$$
{\vec a}^{\it cut}_T({\vec k},t) \buildrel{kt \gg 1}\over=  -
{e^2 \, T^2\over 12 k^2} \,\cos kt \,\left[ \ln 2kt + \gamma_E
\right]\, {\vec a}_T({\vec k},0)
+ {\cal O}\left( \frac{1}{t},\frac{e^4 T^4}{k^4}\right) \;,
$$
where we used~\cite{boyanrgir},
$$
\int_0^{\infty} { dy \over y} \left( 1 - \cos yt \right) \; p(y)
\buildrel{t \to \infty}\over= p(0) \left[ \ln(\mu \, t) + \gamma \right] +
\int_0^{\infty} { dy \over y} \left[p(y) -  p(0) \; \theta(\mu - y )
\right] + {\cal O}\left( {1 \over t } \right) \;.
$$
The pole contribution (\ref{sol:aTpole}) to first order in
${e^2 T^2/k^2}$ reads
$$
{\vec a}^{\it pole}_T({\vec k},t)={\vec a}_T({\vec k},0)\left[\cos kt
-{e^2 \, T^2\over 12\, k^2}\, t \, \sin kt \right] + \mbox{nonsecular~terms}
+ {\cal O}\left({e^4 \, T^4\over k^4}\right) \;.
$$
Collecting pole and cut contributions we find
\begin{eqnarray}
{\vec a}_T({\vec k},t)&=& {\vec a}_T({\vec k},0)\left\{\cos kt -
\frac{e^2 T^2}{12 k^2}\, \left[\left( \ln 2kt
+ \gamma_E \right) \; \cos kt+  t \,\sin kt  \right]\right\} \nonumber\\
&& +\;\mbox{nonsecular~terms} + {\cal O}\left({e^4 T^4\over k^4}\right) \;.
\end{eqnarray}
This result coincides with the perturbative solution (\ref{solord0}),
(\ref{aTsol1s}) and (\ref{masshift}) after imposing the initial
conditions (\ref{condini}) used for the Laplace transform to
Eq.~(\ref{solord0}). That is,
$$
A_\lambda({\vec k})=A^{\ast}_\lambda({\vec k}) \;.
$$
In summary, the Laplace transform solution permits to compute the
secular terms as well as the perturbative method used in sec. II.A2.

\subsection{Quantum kinetics of the photon distribution function}

As mentioned in Sec.~II the first step towards a kinetic description
is to identify the proper degrees of freedom (quasiparticles)
and the corresponding microscopic time scale.
For semihard photons of momenta $eT\ll k\ll T$, it is adequate to
choose the free photons as the quasiparticles with the corresponding
microscopic scale $\sim 1/k$. A kinetic description of the
non-equilibrium evolution of the
distribution function assumes a wide separation between the
microscopic and the relaxation time scales.
In the case under consideration the effective small coupling in the
semihard momentum limit is $e^2T^2/k^2 \ll 1$.
Furthermore {\em assuming} that $e \ll 1$ then in this
regime both the HTL and the perturbative approximation are valid.

We begin by obtaining the photon number operator from the Heisenberg
field operator and its conjugate momentum. Write
\begin{eqnarray*}
{\vec A}_T({\vec x},t)&=&\int\frac{d^3k}{(2\pi)^{3/2}}\;{\vec A}_T({\vec k},t)
\;e^{i{\vec k}\cdot{\vec x}} \;,\\
{\vec P}_T({\vec x},t)&=&\int\frac{d^3k}{(2\pi)^{3/2}}\;{\vec P}_T({\vec k},t)
\;e^{i{\vec k}\cdot{\vec x}} \;,
\end{eqnarray*}
with
\begin{eqnarray*}
{\vec A}_T({\vec k},t)&=&\sum^2_{\lambda=1}\;\sqrt{\frac{1}{2k}}
\;\!\Big[a_\lambda({\vec k},t)\;{\bbox{\cal E}}_\lambda({\vec k})
+a_\lambda^{\dagger}(-{\vec k},t)\;{\bbox{\cal E}}_\lambda(-{\vec
k})\Big] \; ,\\ {\vec P}_T({\vec
k},t)&=&i\sum^2_{\lambda=1}\;\sqrt{\frac{k}{2}}
\;\!\Big[a^{\dagger}_\lambda(-{\vec k},t)\;{\bbox{\cal
E}}_\lambda(-{\vec k}) -a_\lambda({\vec k},t)\;{\bbox{\cal
E}}_\lambda({\vec k})\Big] \; ,
\end{eqnarray*}
where $a_\lambda({\vec k},t)$ [$a^\dagger_\lambda({\vec k},t)$]
is the annihilation (creation) operator that destroys (creates)
a free photon of momentum ${\vec k}$ and polarization $\lambda$ at time $t$.
The polarization-averaged number operator $N_{\gamma}({\vec k},t)$
that counts the semihard photons of momentum ${\vec k}$ is then defined by
\begin{eqnarray*}
N_{\gamma}({\vec k},t)&\equiv&
\frac{1}{2}\sum^2_{\lambda=1}a^\dagger_\lambda({\vec k},t)
a_\lambda({\vec k},t)\\ &=&\frac{1}{4k}\big\{{\vec P}_T(-{\vec
k},t)\cdot{\vec P}_T({\vec k},t) +\;k^2{\vec A}_T(-{\vec
k},t)\cdot{\vec A}_T({\vec k},t)\\ &&+\;ik\big[{\vec A}_T(-{\vec
k},t)\cdot{\vec P}_T({\vec k},t) -\;{\vec P}_T(-{\vec
k},t)\cdot{\vec A}_T({\vec k},t)\big]\big\} \;.
\end{eqnarray*}
The expectation value of this number operator is interpreted as
the number of photons per polarization per unit phase space volume
$$ {n}^{\gamma}_{\vec k}(t)=\langle {N}_\gamma({\vec k},t)\rangle
\equiv (2\pi)^3 \, \frac{dN}{d^3x d^3 k} \;, $$ where $N$ is the
total number of photons per polarization in the plasma. Taking the
time derivative of $N_{\gamma}({\vec k},t)$ and using the
Heisenberg equations of motion, we obtain the following expression
for the time derivative of the expectation value
\begin{equation}
\dot{n}^\gamma_{\vec k}(t)=\lim_{t'\rightarrow t}
\frac{e}{4k}\left(\frac{\partial}{\partial t'}-ik\right)
\int\frac{d^3q}{(2\pi)^{3/2}}\big\langle
\bar{\psi}^-(-{\vec p},t){\bbox\gamma}
\cdot{\vec A}^+_T({\vec k},t')\psi^-({\vec q},t)
\big\rangle + \mbox{c.c.} \;,\label{photon:ndot1}
\end{equation}
where the ``$+$'' (``$-$'') superscripts for the fields
refer to fields defined in the forward (backward) time branch in the CTP
formulation. We have separated the
time arguments and the fields on different branches to be able to
extract the time derivative
from the expectation value.

The nonequilibrium expectation values on the right-hand side (RHS)
of Eq.~(\ref{photon:ndot1}) can be computed perturbatively in powers
of $e$ using the nonequilibrium Feynman rules and real-time propagators.

We assume the initial density matrix is diagonal in the basis of the
photon occupation numbers with
nonequilibrium initial photon populations given by $n^{\gamma}_{\vec k}(t_0)$
and there is no initial photon polarization asymmetry.
Furthermore we assume that the fermions are in thermal equilibrium at
a temperature $T$.

At ${\cal O}(e)$ the RHS of Eq.~(\ref{photon:ndot1}) vanishes identically.
This is a consequence of our choice of initial density matrix
diagonal in the basis of the photon number operator.
To ${\cal O}(e^2)$, we obtain
\begin{equation}
\dot{n}^\gamma_{\vec k}(t)=[1+n^\gamma_{\vec k}(t_0)]
\;\Gamma^<_k(t)-n^\gamma_{\vec k}(t_0)\;\Gamma^>_k(t) \;,\label{photon:ndot2}
\end{equation}
where the time-dependent rates are given by
\begin{eqnarray}
\Gamma^{\mbox{\scriptsize
\raisebox{1.8pt}{\raisebox{1.8pt}{$\scriptscriptstyle>$}
\raisebox{-1pt}{$\scriptscriptstyle\!\!\!\!\!\!<$}}}}_{\vec k}(t)&=&
\int_{-\infty}^{+\infty} d\omega\;{\cal R}^{\mbox{\scriptsize
\raisebox{1.8pt}{\raisebox{1.8pt}{$\scriptscriptstyle>$}
\raisebox{-1pt}{$\scriptscriptstyle\!\!\!\!\!\!<$}}}}_\gamma(\omega,k)\;
\frac{\sin[(\omega-k)(t-t_0)]}{\omega-k},\nonumber\\
{\cal R}^<_\gamma(\omega,k)&=&\frac{e^2}{k}\int\frac{d^3q}{(2\pi)^{3}}
\bigg[\big[1-(\hat{\vec k}\cdot\hat{\vec p})
(\hat{\vec k}\cdot\hat{\vec q})\big]
\Big\{n_F(p)[1-n_F(q)]\delta(\omega-p+q) \nonumber\\
&&+\;[1-n_F(p)]n_F(q)\delta(\omega+p-q)\Big\}
+\big[1+(\hat{\vec k}\cdot\hat{\vec p})
(\hat{\vec k}\cdot\hat{\vec q})\big]\nonumber\\
&&\times\Big\{n_F(p)n_F(q) \delta(\omega-p-q)
+[1-n_F(p)]\;[1-n_F(q)]\delta(\omega+p+q)\Big\}\bigg] \;,
\label{Rgamma1}
\end{eqnarray}
with ${\vec p}={\vec k}+{\vec q}$ and ${\cal R}^>_\gamma(\omega,k)$ is
obtained from  $ {\cal R}^<_\gamma(\omega,k) $ through the replacement
$n_F\leftrightarrow 1-n_F$.

A comment here is in order. As explained above we are focusing on
the leading HTL approximation, consequently,  in obtaining ${\cal
R}^{\mbox{\scriptsize
\raisebox{1.8pt}{\raisebox{1.8pt}{$\scriptscriptstyle>$}
\raisebox{-1pt}{$\scriptscriptstyle\!\!\!\!\!\!<$}}}}_\gamma(\omega,k)$
we use the {\em free} real-time fermion propagators which
correspond to the {\em hard} part of the fermion loop momentum. In
general there are contributions from the soft fermion loop
momentum region  which will require to use the HTL-resummed
fermion propagators~\cite{htl,rob2,rob3} for consistency. We will
postpone the detailed study of the contribution from soft loop
momentum  to a forthcoming article, focusing here on the
comparison between the lowest order calculation {\em in real time}
and the result available in the literature for the photon
production {\em rate}  that include HTL corrections.

Since the fermions are in thermal equilibrium,
the Kubo-Martin-Schwinger (KMS) condition holds:
\begin{equation}
{\cal R}^>_\gamma(\omega,k)=e^{\beta\omega} \;
{\cal R}^<_\gamma(\omega,k)\label{KMScond} \;,
\end{equation}
where $\beta=1/T$.
It is apparent to see that ${\cal R}^{<(>)}_\gamma(\omega,k)$ has a physical
interpretation in terms of the {\em off-shell} photon production (absorption)
processes in the plasma.
The first two terms in ${\cal R}^<_\gamma(\omega,k)$
describe the process that a fermion (or an anti-fermion) emits a
photon, i.e, bremsstrahlung from the fermions in the medium,
the third term describes annihilation of a fermion pair into a photon,
and the fourth terms describes creation of a photon and a fermion pair
out of the vacuum.
The corresponding terms in ${\cal R}^>_\gamma(\omega,k)$ describe the
inverse processes.

As argued, for semihard photons of momenta $eT\ll k\ll T$,
the leading contribution of ${\cal R}^{\mbox{\scriptsize
\raisebox{1.8pt}{\raisebox{1.8pt}{$\scriptscriptstyle>$}
\raisebox{-1pt}{$\scriptscriptstyle\!\!\!\!\!\!<$}}}}_\gamma(\omega,k)$
arises from the hard loop momenta $q\sim k$.
A detailed analysis of ${\cal R}^<_\gamma(\omega,k)$ along the
familiar lines in the
hard thermal loop program\cite{htl,rob2,rob3,lebellac} shows that
in the HTL approximation ($q\gg k$)
\begin{equation}
{\cal R}^<_\gamma(\omega,k)|_{\rm HTL}=\frac{e^2T^3}{12k^2}
\left(1-\frac{\omega^2}{k^2}\right)
\;\theta(k^2-\omega^2).\label{Rgamma2}
\end{equation}
Thus we recognize that in the HTL limit  ${\cal R}^{<(>)}_\gamma(\omega,k)$
is completely determined to the off-shell Landau damping process
in which a hard fermion in the plasma emits (absorbs) a semihard photon,
i.e, bremsstrahlung from the fermions in the medium.

\subsubsection*{Dynamical Renormalization Group and the emergence of
detailed balance}

We now turn to the kinetics of semihard photons. To obtain a kinetic
equation from Eq.~(\ref{photon:ndot2}), we implement the dynamical
renormalization group resummation as introduced in
Refs.~\cite{boyanrgir,boyanrgk}.
Direct integration with the initial condition yields
\begin{eqnarray}
n^\gamma_{\vec k}(t)&=&n^\gamma_{\vec k}(t_0)+\left[1+n^\gamma_{\vec
k}(t_0)\right]
\int^t_{t_0}dt'\;\Gamma^<_k(t')
-\;n^\gamma_{\vec k}(t_0)\int^t_{t_0}dt'\;\Gamma^>_k(t')\; .\label{ngamma2}
\end{eqnarray}
The integrals that appear in the above expression:
\begin{equation}
\int^t_{t_0}dt'\; \Gamma^{\mbox{\scriptsize
\raisebox{1.8pt}{\raisebox{1.8pt}{$\scriptscriptstyle>$}
\raisebox{-1pt}{$\scriptscriptstyle\!\!\!\!\!\!<$}}}}_{\vec k}(t') =
\int_{-\infty}^{+\infty} d\omega\;{\cal R}^{\mbox{\scriptsize
\raisebox{1.8pt}{\raisebox{1.8pt}{$\scriptscriptstyle>$}
\raisebox{-1pt}{$\scriptscriptstyle\!\!\!\!\!\!<$}}}}_\gamma(\omega,k)\;
\frac{1-\cos[(\omega-k)(t-t_0)]}{(\omega-k)^2} \;,
\label{timedeprates}
\end{equation}
are dominated, in the long time limit, by the regions of
$\omega$ for which the denominator is resonant, i.e, $\omega\approx k$.
The time dependence in the above integral along with the resonant
denominator is the familiar form that
leads to Fermi's golden rule in elementary time dependent perturbation
theory. In the long time limit $ t-t_0 \gg 1/k$ Fermi's golden rule
approximates the above integrals by
$$
\int^t_{t_0}dt'\; \Gamma^{\mbox{\scriptsize
\raisebox{1.8pt}{\raisebox{1.8pt}{$\scriptscriptstyle>$}
\raisebox{-1pt}{$\scriptscriptstyle\!\!\!\!\!\!<$}}}}_{\vec k}(t')\approx
\frac{\pi}{2} (t-t_0)  {\cal R}^{\mbox{\scriptsize
\raisebox{1.8pt}{\raisebox{1.8pt}{$\scriptscriptstyle>$}
\raisebox{-1pt}{$\scriptscriptstyle\!\!\!\!\!\!<$}}}}_\gamma(\omega=k,k)
\equiv 0 \; .
$$
Therefore the {\em rate} of photon production, i.e, the coefficient of the
linear time dependence vanishes at this order because of the
vanishing of the imaginary part of the photon self-energy on the
photon mass shell. However the use of Fermi's golden rule,
which is the usual approach to extract (time-independent) rates,
misses the important off-shell effects associated
with finite lifetime processes. These can be understood explicitly by
using the first of formulae (\ref{formulas1}).
We find that for $ t-t_0 \gg 1/k $
\begin{eqnarray}
\int^t_{t_0} dt'\; \Gamma^<_k(t') & \buildrel{k(t-t_0) \gg 1}\over= & \frac{e^2
T^3}{6k^3}\left\{\ln\left[2k(t-t_0) \right]+\gamma_E-1\right\}+{\cal
O}\left(\frac{1}{k(t-t_0)}\right) \nonumber \\
\int^t_{t_0}dt'\;\Gamma^>_k(t') & \buildrel{k(t-t_0) \gg 1}\over = &
e^{\beta k} \int^t_{t_0}dt'\;\Gamma^<_k(t') \;. \label{detbalance}
\end{eqnarray}
The second line of this equation displays the condition for
{\em detailed balance} and holds for time scales $t-t_0 \gg 1/k$.
We emphasize that this condition is a consequence of the fact that the
region $\omega \approx k$ (i.e, the resonant denominator) dominates the
long time behavior of the integrals in the time-dependent rates
(\ref{timedeprates}) and the KMS condition Eq.~(\ref{KMScond})
holds under the assumption that the fermions are in thermal equilibrium.

Thus we obtain one of the important results of this article {\em the
time-dependent rates obey a detailed balance} and
that detailed balance emerges in the intermediate asymptotic regime
$ t-t_0 \gg 1/k$ when the secular terms dominate the
long time behavior. This is a noteworthy result because it clearly
states that  detailed balance emerges on {\em microscopic} time scales
$ t-t_0 > 1/k$ at which the secular terms dominate the integrals but
for which a perturbative expansion is still valid. Detailed balance
then  guarantees the existence of an asymptotic equilibrium solution
which is reached on time scales of the order of the relaxation time.

The (logarithmic) secular terms in the time-dependent rates
(\ref{detbalance}) can now be resummed using the
dynamical renormalization group method~\cite{boyanrgir,boyanrgk} by
introducing a multiplicative renormalization of the distribution function
\begin{eqnarray*}
n^\gamma_{\vec k}(t_0)&=&{\cal Z}_k(t_0,\tau) \; n^\gamma_{\vec
k}(\tau)\; ,\\
{\cal Z}_k(t_0,\tau)&=&1+e^2 z_k^{(1)}(t_0,\tau)+{\cal O}(e^4)\; ,
\end{eqnarray*}
thus rewriting Eq.~(\ref{ngamma2}) consistently to ${\cal O}(e^2)$  as
\begin{eqnarray}
n^\gamma_{\vec k}(t)&=&n^\gamma_{\vec k}(\tau)+
e^2 z_k^{(1)}(t_0,\tau) \;n^\gamma_{\vec k}(\tau)+
\left[1+n^\gamma_{\vec k}(\tau)\right]\int^t_{t_0}dt' \;\Gamma^<_k(t')
-n^\gamma_{\vec k}(\tau)  \int^t_{t_0}dt' \;\Gamma^>_k(t')+{\cal
O}(e^4) \;. \nonumber
\end{eqnarray}
The renormalization coefficient $z^{(1)}_k(t_0,\tau) $ is
chosen to cancel the secular divergence at a time scale $t=\tau$.
Thus to ${\cal O}(e^2)$, the choice
\begin{eqnarray*}
e^2 z_k^{(1)}(t,\tau)\;n^\gamma_{\vec k}(\tau)
&=&-\left[1+n^\gamma_{\vec k}(\tau)\right] \int^\tau_{t_0}dt'\;\Gamma^<_k(t')
+\;n^\gamma_{\vec k}(\tau)\int^\tau_{t_0}dt'\;\Gamma^>_k(t') \;,
\end{eqnarray*}
leads to an {\em improved} perturbative solution in terms of the ``updated''
occupation number $n^\gamma_{\vec k}(\tau)$,
\begin{eqnarray}
n^\gamma_{\vec k}(t)&=&n^\gamma_{\vec k}(\tau)+
\left[1+n^\gamma_{\vec k}(\tau)\right]\int^t_{\tau}dt'\;\Gamma^<_k(t')
-n^\gamma_{\vec k}(\tau) \int^t_{\tau}dt'\;\Gamma^>_k(t')+
{\cal O}(e^4) \;, \nonumber
\end{eqnarray}
which is valid for large times $t \gg t_0$ provided
that $\tau$ is chosen arbitrarily close to $t$.
A change in the time scale $\tau$ in the integrals is compensated by a
change of the $n^\gamma_{\vec k}(\tau)$ in such
a manner that $n^\gamma_{\vec k}(t)$ does not depend on the arbitrary
scale $\tau$. This independence leads to the dynamical
renormalization group equation~\cite{boyanrgir,boyanrgk} which
consistently to ${\cal O}(e^2)$ is given by
\begin{equation}
\frac{d}{d\tau}{n}^\gamma_{\vec k}(\tau)=[1+n^\gamma_{\vec k}(\tau)]
\;\Gamma^{<}_k(\tau)-
n^\gamma_{\vec k}(\tau)\; \Gamma^{>}_k(\tau)+{\cal O}(e^4)
\;.\label{photon:drg}
\end{equation}
Choosing $\tau$ to coincide with $t$ in Eq.~(\ref{photon:drg}),
we obtain the {\em quantum kinetic equation} to order $e^2$  given by
\begin{equation}
\dot{n}^\gamma_{\vec k}(t)=[1+n^\gamma_{\vec k}(t)]\;\Gamma^{<}_k(t)-
n^\gamma_{\vec k}(t)\;\Gamma^{>}_k(t) \; .\label{photon:keq}
\end{equation}
For intermediate asymptotic times $ t-t_0 \gg 1/k $
at which the logarithmic secular terms
dominate the integrals for the time-dependent rates and
{\em detailed balance} emerges, we find
\begin{eqnarray}
\Gamma^{<}_k(t)&\buildrel{k(t-t_0) \gg
1}\over=&\frac{e^2T^3}{6k^3}\frac{1}{t-t_0}
\left[1+{\cal O}\left(\frac{1}{k(t-t_0)}\right)\right],\nonumber \\
\Gamma^{>}_k(t)&\buildrel{k(t-t_0) \gg
1}\over=&\frac{e^2T^3}{6k^3}\frac{e^{\beta k}}{t-t_0}
\left[1+{\cal O}\left(\frac{1}{k(t-t_0)}\right)\right] \;,\nonumber
\end{eqnarray}
where the detailed balance relation is explicitly displayed and the
terms being neglected are oscillatory on time scales $\sim 1/k$ and
fall off faster.

The detailed balance relation between the time-dependent rates
(\ref{photon:keq}) guarantees the existence of an {\em equilibrium}
solution of the kinetic equation with the occupation number
$n^\gamma_{\vec k}(t=\infty) = (e^{\beta k}-1)^{-1}$.

The full solution of this kinetic equation is given by
\begin{eqnarray}
n^\gamma_{\vec k}(t) & = &  n^\gamma_{\vec k}(t_0) \; e^{-\int^t_{t_0}dt'
\gamma_k(t')} + e^{-\int^t_{t_0}dt'\gamma_k(t')}
\int^t_{t_0} dt'\;\Gamma^{<}_k(t')\; e^{\int^{t'}_{t_0}dt''\gamma_k(t'')},
\label{fullsolkin} \\
\gamma_k(t) & = & \Gamma^{>}_k(t)-\Gamma^{<}_k(t) \;. \nonumber
\end{eqnarray}
However, in order to understand the relaxation to equilibrium of the
distribution function, we now focus on the linearized
relaxation time approximation which describes the approach
to equilibrium of the distribution function of {\em one}
{\vec k}-mode which has been displaced slightly off
equilibrium while all other modes (and the fermions) are in equilibrium.
Thus writing $n^\gamma_{\vec k}(t_0)=n_B(k)+\delta n^\gamma_{\vec k}(t_0)$
with $n_B(k)$ the Bose-Einstein distribution function, which as shown
above is an equilibrium solution as a consequence
of detailed balance (\ref{detbalance}), we obtain
$$
\delta\dot{n}^\gamma_{\vec k}(t)=
-\delta n^\gamma_{\vec k}(t) \;(e^{\beta k}-1) \;\Gamma^{<}_k(t)\;.
$$
For semihard photons of momentum $eT\ll k\ll T$, we can simply replace
$e^{\beta k}-1$ by $k/T$ and upon integration we find
\begin{equation}
\delta n^\gamma_{\vec k}(t)=\delta n^\gamma_{\vec k}(t_0)
\left(\frac{t-t_0}{\tau_0}\right)^{-\frac{e^2 T^2}{6k^2}}
\quad\mbox{ for}\quad k(t-t_0)\gg 1 \; ,\label{rta}
\end{equation}
where $\tau_0\sim 1/k$.

A noteworthy feature of Eq.~(\ref{rta}) is that the relaxation of the
distribution function for semihard photons
in the linearized approximation is governed by a power law
with an anomalous exponent rather than by exponential
damping. This result is similar to that found
in scalar quantum electrodynamics in the Markovian
approximation~\cite{boyanhtl}.
Comparing Eq.~(\ref{rta}) and Eq.~(\ref{finalsol}), we clearly see that the
anomalous exponent for the relaxation of the photon distribution
function in the linearized approximation is {\em twice} that for the
linear relaxation of the mean field (\ref{finalsol}).

This relationship is well known in the case of exponential relaxation
but our analysis with the dynamical renormalization
group reveals it to be a more robust feature, applying just as well to
power law relaxation.

Our approach to derive kinetic equations in field theory is
different from the one often used in the literature which involves
a Wigner transform and assumptions about the separation of fast
and slow variables~\cite{geiger,heinz,mrow,daniel,niegawa}. Our
work differs in many important respects from these other
formulations in that it reveals clearly the dynamics of off-shell
effects associated with non-exponential relaxation. This aspects
will acquire phenomenological relevance in our study of photon
production by off-shell effects in a plasma with a finite lifetime
in the next section.

\subsection{Photon production}

An important and phenomenologically relevant byproduct of the
study of the kinetics of the photon distribution function is to
give an assessment of photon production via {\em off-shell effects}
by the lightest (up and down) quarks from the quark-gluon
plasma in which quarks and gluons (but not photons) are in thermal
equilibrium. To the order under consideration the only
contribution to the photon self-energy is a fermion loop and
although we have computed it assuming these fermions to be
electrons, we can use this result to study photon production from
thermalized quarks by simply accounting for the electromagnetic
charges of the up and down quarks and for the number of colors.
To this order ${\cal O}(\alpha)$ there are no gluon contributions
to the photon self-energy and therefore to photon production and
the rate for photon production obtained from the imaginary part of the
photon self-energy evaluated on the photon mass shell vanishes.
This can be seen directly from Eqs.~(\ref{Rgamma2}) and (\ref{timedeprates})
which show explicitly that the usual (time-independent) rate obtained
from Fermi's golden rule and determined by the imaginary part of the
photon self-energy on the photon mass shell vanishes.
The underlying assumption that motivates the use of Fermi's golden rule
is that the lifetime of the system is much larger than typical
observation times and the limit of $t\rightarrow \infty$ can
be taken replacing the oscillatory factors with resonant denominators
in Eq.~(\ref{timedeprates}) by a delta function.
This obviously extracts the leading time dependence but
ignores corrections arising from the finite lifetime of the system
under consideration. A quark-gluon plasma
formed in an ultrarelativistic heavy ion collision is estimated to
have a lifetime $\lesssim 50\;\mbox{fm}/c$ at RHIC energies thus
photons are produced by a source of a finite lifetime and off-shell
effects could lead to considerable photon production during the lifetime
of the QGP perhaps comparable to those extracted from the usual Fermi's
golden rule results for infinite lifetime.
The focus of this section is precisely the study of this possibility .

From the expression for the number of photons per phase space
volume given by Eq.~(\ref{fullsolkin}) we can obtain the number of
photons produced at a given time $t$ per unit phase space from an
initial (photon) vacuum state, by setting $n^{\gamma}_{\vec
k}(t_0)=0$ and including the charge of the up and down quarks and
the number of colors. Assuming that photons do not thermalize in
the quark-gluon plasma, i.e, that their mean free path is much
longer than the size of the plasma, we neglect the exponentials in
the second term of  Eq.~(\ref{fullsolkin}) which are responsible
for building the photon population up to the equilibrium value. To
lowest order in the electromagnetic fine structure constant we
find from Eq.~(\ref{fullsolkin}) the total number of semihard
photons at time $t$ per invariant phase space volume summed over
the two polarizations  given by
\begin{equation}\label{alpha}
k \int^t_{t_0}dt'\frac{dN_{\rm tot}(t')}{d^3x\,d^3 k\,dt'} \equiv
\frac{2k}{(2\pi)^3} n^{\gamma}_{\vec k}(t) \buildrel{k(t-t_0) \gg 1}\over=
\frac{5\alpha T^3}{18 \pi^2 k^2}\;\left[\ln 2k(t-t_0)+\gamma_E-1\right]+
{\cal O}\left[(t-t_0)^{-1}\right].
\end{equation}
This is a noteworthy result and one of the main points of this article, we find
photon production to lowest (one-loop) order arising solely from
{\em off-shell} effects:
($q\rightarrow q\gamma$ and $\bar{q}\rightarrow\bar{q}\gamma$) and
annihilation of quarks ($q\bar{q} \rightarrow\gamma$).
In the HTL limit, valid for semihard photons, the leading
contribution arises from Landau damping, i.e,  from
quarks in the medium.

In the usual approach~\cite{kapusta,baier} the photon production rate
is obtained from the imaginary part of the photon self-energy {\em on-shell}
and receives contributions only at order ${\cal O}(\alpha\,\alpha_s)$
and beyond.  The lowest order contribution corresponds to a quark loop
with a self-energy insertion from gluons as well as electromagnetic
vertex correction by gluons.
As a result the imaginary part of the photon self-energy on
the mass shell arises from Compton scattering
($qg\rightarrow q\gamma$ and $\bar{q}g\rightarrow \bar{q}\gamma$)
and pair-annihilation ($q\bar{q}\rightarrow g\gamma$)~\cite{kapusta,baier}.
The lowest order ${\cal O}(\alpha\,\alpha_s)$ contribution displays an infrared
divergence for massless quarks that is screened by the HTL resummation
leading to the result quoted in Ref.~\cite{kapusta,baier}
for the {\em hard} photon production rate~\cite{kapusta}:
\footnote{We follow Kapusta {\it et al.}~\cite{kapusta}
in adding 1 to the argument of the logarithm so
as to make it agree with the numerical result for $k\sim T$. For
details, see discussion below Eq.~(41) in Ref~\cite{kapusta}.}
$$
k\frac{dN_{\rm tot}(t)}{d^3x\,d^3k \,dt}\bigg|_{\text{on-shell}}=
\frac{5}{18 \pi^2} \,{\alpha \,\alpha_s}\,T^2 \, e^{-k/T}\,
\ln\left(\frac{2.912}{g^2}\frac{k}{T}+1\right) \;,
$$
where $\alpha_s=g^2/4\pi$ with $g$ the QCD coupling constant.
The total number of photons produced per (invariant) phase space at
time $t$ is then given by
\begin{equation}
k\int^t_{t_0}dt'\frac{dN_{\rm
tot}(t')}{d^3x\,d^3k\,dt'}\bigg|_{\text{on-shell}}=
\frac{5}{18 \pi^2} \, \alpha \, \alpha_s\, T^2 \,(t-t_0)\,e^{-k/T}\,
\ln\left(\frac{2.912}{g^2}\frac{k}{T}+1\right) \;,\label{alphaalphas}
\end{equation}
which can obviously be interpreted as a {\em linear}
secular term in the photon distribution function, which is
the usual situation corresponding to exponential relaxation of the photon
distribution function~\cite{boyanrgk}
as well as for the photon mean field~\cite{boyanrgir}.

\subsubsection*{Comparison between on- and off-shell photon production}

The important question to address is how do the on- and off-shell
contributions to photon production compare? Obviously the answer
to this question depends on the time scale involved, eventually at
long time scales the linear growth of the photon number with time
from the on-shell contribution will overwhelm the logarithmic
growth from the off-shell, but at early times the off-shell will dominate.
Thus a crossover regime is expected from the off-shell
to the on-shell dominance to photon production.
This crossover time scale will depend on the temperature
and the energy of the photons as well as on the numerical value of
the electromagnetic and strong couplings.

There are two important time scales that are of phenomenological
relevance to photon production in a quark-gluon plasma.
Firstly the assumption that quarks are in thermal
equilibrium restricts the earliest time scale to be somewhat
larger than about $1-3\;\mbox{fm}/c$ which is the estimate for
quark thermalization~\cite{qgp1,qgp2}.
Secondly the lifetime of the QGP phase is estimated to be somewhere in
the range $10-50\;\mbox{fm}/c$.
Thus the relevant time scale for comparison between the off- and on-shell
contributions is $1-3\;\mbox{fm}/c\lesssim t-t_0\lesssim 10-50\;\mbox{fm}/c$.

\begin{center}
\begin{figure}
\epsfig{file=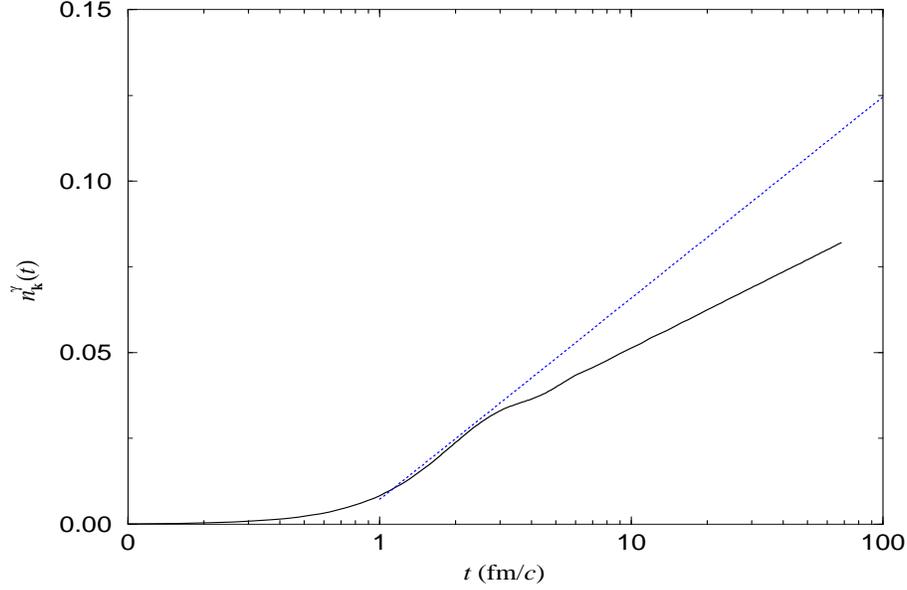,width=12.0cm,height=8.0cm}
\vspace{.1in} \caption{Off-shell contribution to $n^{\gamma}_{\vec
k}(t)$ vs $t$ for hard photons $k \sim T$, $T=200\;\mbox{ MeV}$
and $\alpha=1/137$. The solid line is the numerical evaluation
using the full time dependent rates, the dashed line is the HTL
approximation. \label{fig:qcdphotonprod}}
\end{figure}
\end{center}

\begin{center}
\begin{figure}
\epsfig{file=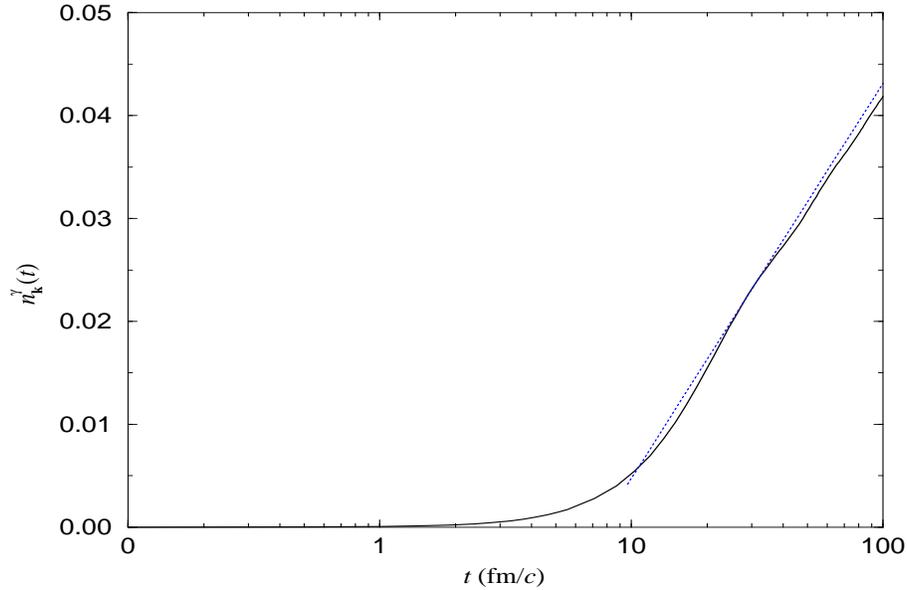,width=12.0cm,height=8.0cm}
\vspace{.1in} \caption{The number of semihard photons is plotted
as a function of time for $e=0.01$, $T=200\;\mbox{MeV}$ and
$eT/k=0.1$. The solid line is the numerical evaluation using the
full time dependent rates, the dashed line is the HTL
approximation. \label{fig:photonprod}}
\end{figure}
\end{center}

Before we begin the comparison, we point out that
Eqs.~(\ref{alpha}) and (\ref{alphaalphas}) have different
regimes of validity in photon momenta.
Whereas  Eq.~(\ref{alpha}) is valid in the semihard regime $eT \ll k \ll T$
where the HTL approximation is reliable,
Eq.~(\ref{alphaalphas}) is valid in the hard regime $k\gtrsim T$~\cite{kapusta,baier}.
We have studied numerically the off-shell
contribution to photon production for hard momentum $k \sim T $
directly in terms of the time-dependent rates given by
Eqs.~(\ref{photon:ndot2}) and (\ref{Rgamma1}) without using the HTL
approximation and compared it to the HTL approximation for
$1\;\mbox{fm}/c \leq t \leq 100 \; \mbox{fm}/c$ (with $ t_0=0 $).
The result is depicted in Fig.~\ref{fig:qcdphotonprod}, which displays
clearly a logarithmic time dependence in both cases for
$t > 1-2\;\mbox{fm}/c$ but with different slopes, a consequence of the
different temperature and momentum dependence of the coefficients
of the logarithmic time dependence.
The HTL approximation overestimates the total number of
photons by at most a factor two in the relevant regime
$1\;\mbox{fm}/c \leq t \leq 100 \;\mbox{fm}/c$ for hard photons.
The reliability of the HTL approximation (\ref{alpha}) in the semihard momentum
region is confirmed by Fig.~\ref{fig:photonprod}, which compares the
number of photons obtained from the numerical evaluation with the full
time-dependent rates and the result from the HTL approximation in the
weak coupling and semihard momentum region.

We can now compare the off-shell contribution to that of
Refs.~\cite{kapusta,baier}, which is a result of on-shell processes
and valid in the hard momentum limit. For this we focus on the
relevant scenario of a quark-gluon plasma with thermalized quarks
at a temperature $T\sim 200\;\mbox{MeV}$ and of lifetime $t\sim
10-20\;\mbox{fm}/c$. These are approximately the temperature and
temporal scales expected to be reached at RHIC. The value of
$\alpha_s$ at this temperature is not know with much certainty but
expected to be $\lesssim 1 $. Following~\cite{kapusta} we choose
$\alpha_s=0.4$ which corresponds to $g^2=5$, we find that for hard
photons of momenta $k\sim T$ the ratio of photon produced by
off-shell processes is {\em comparable} to that produced by on-shell processes.

\begin{center}
\begin{figure}
\epsfig{file=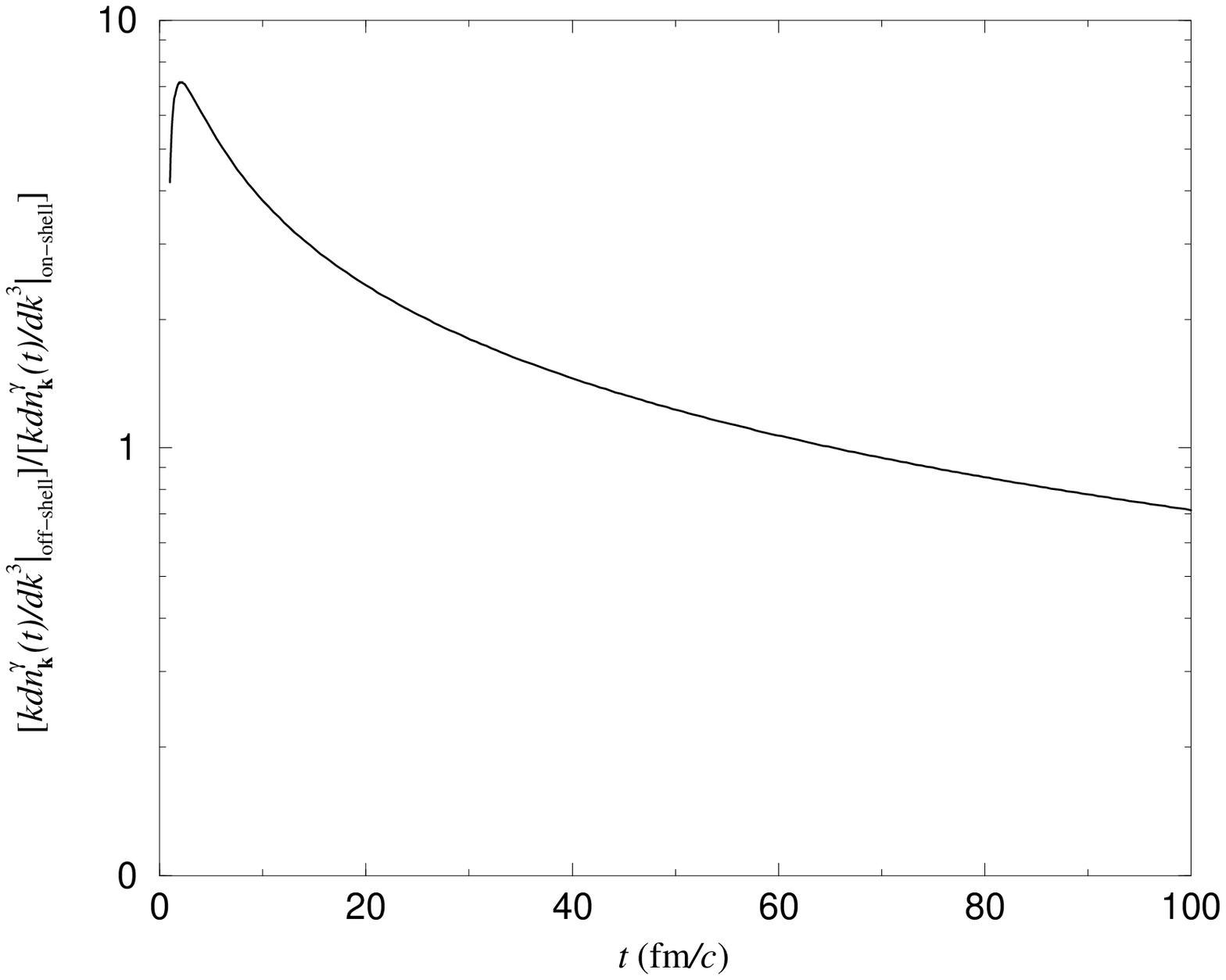,width=12.0cm,height=8.0cm}
\vspace{.1in} \caption{The ratio of the number of hard photons
produced by off-shell process ($q\rightarrow q\gamma$ and
$\bar{q}\rightarrow\bar{q}\gamma$) to that produced by on-shell
processes ($qg\rightarrow q\gamma$ and $q\bar{q}\rightarrow
g\gamma$) as a function of QGP lifetime $t$ for $\alpha=1/137$,
$\alpha_s=0.4$ and $T=200\;{\rm MeV}$. \label{fig:photonratio}}
\end{figure}
\end{center}

Fig.~\ref{fig:photonratio} depicts the ratio of the number of hard
($k\sim T$)  photons produced by off-shell processes in the HTL
approximation to that from on-shell processes in the regime of
lifetimes for a quark-gluon plasma phase
$5\;\mbox{fm}/c<t<100\;\mbox{fm}/c$ (here we set $t_0=0$).
Accounting for the overestimate of the HTL approximation for hard photons,
we then conclude that off-shell processes such as  from
quarks in the medium (Landau damping) are just {\em as important as}
on-shell processes. This is another of the important results of
this work and that cannot be obtained from the usual approach to
photon production based on the computation of a time-independent rate.

Furthermore, an important point worth emphasizing is that whereas the on-shell
calculation displays infrared divergences at lowest order, the
calculation directly in
real time is infrared finite because time acts as an infrared
cutoff. Thus a calculation of
photon production directly in real time does not  require an HTL
improvement of the fermion
propagators to cutoff an infrared divergence. However, the region of
soft loop momentum does
require the HTL resummation  of one of the fermion propagators and
will be studied in a future article.

A thorough study of off-shell effects and their contribution to photon
production in a wider range of temperatures and momenta including
screening corrections to the internal quark lines and the study of
potential observables will be the subject of a longer study which we
think is worthwhile on its own and on which we expect to report soon.


\section{Hard fermions out of equilibrium}

To provide a complete picture of relaxation and non-equilibrium
aspects of a hot QED plasma in real time, we now focus
on a detailed study of relaxation of fermionic mean fields (as induced
by an adiabatically switched-on Grassmann source) as
well as the quantum kinetics of the fermion distribution function. In
this section we assume that the photons are in thermal
equilibrium, and since we work to leading order in the HTL
approximation we can translate the results {\em vis \`a vis} to
the case of equilibrated gluons. In particular we seek to study the
possibility of anomalous relaxation as a result
of the emission and absorption of magnetic photons.
In Refs.~\cite{blaizotBN,taka} it was found that the relaxation of
fermionic excitations is anomalous and not exponential as a result of
the emission and absorption of magnetic photons
that are only dynamically screened by Landau damping. The study of the
real-time relaxation of the fermionic mean fields
in these references was cast in terms of the Bloch-Nordsieck
approximation which replaces the  gamma matrices by the
classical velocity of the fermion. In Ref.~\cite{boyanrgir} the
relaxation of a charged scalar mean field as well
as the quantum kinetics of the distribution function of charged
scalars in scalar electrodynamics were studied
using the dynamical renormalization group, both the charged scalar
mean field and the distribution function of charged
particles reveal anomalous non-exponential relaxation as a consequence
of emission and absorption of soft magnetic photons.
While electric photons (plasmons) are screened by a Debye mass
which cuts off their infrared contribution, magnetic photons are only
dynamically screened by Landau damping and
their emission and absorption dominates the infrared behavior of the
fermion propagator.

While the dynamical renormalization group has been implemented in scalar
theories it has not yet been applied to fermionic
theories. Thus the purpose of this section is twofold,
(i) to implement the dynamical renormalization group to
study the relaxation and kinetics of fermions with a
detailed discussion of the technical differences with the
bosonic case and (ii) to focus on the real-time manifestation
of the infrared singularities associated with soft magnetic photons.

\subsection{Relaxation of the fermionic mean field}

The equation of motion for a fermionic mean field is obtained by
following the strategy described in section II. We begin
by writing the fermionic  field as
\begin{eqnarray*}
\Psi^\pm({\vec x},t)&=&\psi({\vec x},t)+\chi^\pm({\vec x},t),\\
\psi({\vec x},t)&=&\langle\Psi^\pm({\vec x},t)\rangle \;.
\end{eqnarray*}
Then using the tadpole method~\cite{tadpole}, with an external
Grassmann source that is adiabatically switched-on from $t=-\infty$
and switched-off at $t=0$, we  find the Dirac equation for the
spatial Fourier transform of the fermion mean field
for $t>0$  given by
\begin{equation}
\left(i\gamma_0\frac{\partial}{\partial t}
- {\bbox\gamma}\cdot{\vec k}\right)\psi({\vec k},t) -
\int^{t}_{-\infty}dt'\; \Sigma({\vec k},t-t')\;\psi({\vec
k},t')=0 \;,\label{diraceq1}
\end{equation}
where $\Sigma({\vec k},t-t')$ is the retarded fermion self-energy.
A comment here is in order. To facilitate the study and maintain
notational simplicity, in obtaining the equation we neglect
the contribution from the instantaneous Coulomb interaction which
is irrelevant to the relaxation of the mean field and only results
in a perturbative frequency shift.

As noted above the relaxation of hard fermions is dominated
by the soft photon contributions, thus in a perturbative expansion
one needs to use the HTL-resummed photon propagators to account
for the screening effects in the medium.
To one-loop order but with the HTL-resummed photon propagators
given in the Appendix, $\Sigma({\vec k},t-t')$ reads
\begin{eqnarray}
\Sigma({\vec k},t-t')&=&\int_{-\infty}^{+\infty}
d\omega \Big[ -i\,\gamma_0\,\rho_1(\omega,{\vec k})\cos[\omega(t-t')]
+\;{\bbox\gamma}\cdot{\vec \hat{k}}\,\rho_2(\omega,{\vec k})
\sin[\omega(t-t')]\Big] \;,
\end{eqnarray}
where the spectral densities,
\begin{eqnarray}
\rho_1(\omega,{\vec k})&=&e^2\int\frac{d^3 q}{(2\pi)^3}
\int_{-\infty}^{+\infty} dq_0\;[1+n_B(q_0)-n_F(p)]\nonumber\\
&&\times\Big[\tilde{\rho}_T(q_0,q)+\frac{1}{2}\;\tilde{\rho}_L(q_0,q)\Big]
[\delta(\omega-p-q_0)+ \delta(\omega+p+q_0)],\label{rho1}\\
\rho_2(\omega,{\vec k})&=&e^2\int\frac{d^3 q}{(2\pi)^3}
\int_{-\infty}^{+\infty} dq_0\;[1+n_B(q_0)-n_F(p)]\nonumber\\
&&\times\Big[({\vec\hat k}\cdot{\vec\hat q})
({\vec\hat p}\cdot{\vec\hat q})\;\tilde{\rho}_T(q_0,q)
-\frac{{\vec\hat k}\cdot{\vec\hat p}}{2}\;\tilde{\rho}_L(q_0,q)\Big]
\;[\delta(\omega-p-q_0)- \delta(\omega+p+q_0)] \;,\label{rho2}
\end{eqnarray}
with ${\vec p}={\vec k}-{\vec q}$.
Here $\tilde{\rho}_T(q_0,q)$ is the HTL-resummed spectral density for
transverse photon propagator defined in Eq.~(\ref{tilderhot}) and
$\tilde{\rho}_L(q_0,q)$ is the HTL-resummed spectral density for
longitudinal photon propagator
\begin{eqnarray}
\tilde{\rho}_{L}(q_0,q)&=&\mbox{sgn}(q_0)\,Z_{L}(q)\,
\delta[q^2_0-\omega^2_{L}(q)]
+\;\beta_{L}(q_0,q)\;\theta(q^2-q_0^2)\;, \nonumber\\
\beta_{L}(q_0,q)&=&
\frac{\frac{e^2T^2}{6}\frac{q_0}{q}}{\left[q^2+
\frac{e^2T^2}{6}\left(2-\frac{q_0}{q}\ln\frac{q+q_0}{q-q_0}\right)\right]^2+
\big[\frac{\pi e^2T^2}{6}\frac{q_0}{q}\big]^2}\;,\label{rholong}
\end{eqnarray}
where $\omega_{L}(q)$ is the plasmon (longitudinal photon) pole
and $Z_{L}(q)$ is the corresponding residue~\cite{lebellac}.
It is worth pointing out that $\rho_{1(2)}(\omega,{\vec k})$
is an even (odd) function of $\omega$, a property that will be useful
in the following analysis. Furthermore, to establish
a connection with results in the literature, it proves convenient to
introduce the Laplace transform of the retarded
self-energy $\tilde{\Sigma}(s,\vec k)$ just as in
Eq.~(\ref{laplatrans}) and its analytic continuation
$\Sigma(\omega,\vec k)$ as in Eq.~(\ref{pi:anacont}) which is given by
$$
\Sigma(\omega,{\vec k}) = \int^{+\infty}_{-\infty}
\frac{d\nu}{\nu-\omega-i0^+}
\left[-\gamma_0\rho_1(\nu,{\vec k})+{\bbox\gamma}\cdot{\vec\hat{k}}
\;\rho_2(\nu,{\vec k})\right] \;.
$$

Following the same strategy in the study of the photon mean field,
we define $\sigma({\vec k},t-t')$ as
$$
\Sigma({\vec k},t-t')=\frac{\partial}{\partial t'}\sigma({\vec
k},t-t') \;,
$$
and rewrite Eq.~(\ref{diraceq1}) as an initial value problem
\begin{eqnarray}
\left(i\gamma_0\frac{\partial}{\partial t}
- {\bbox\gamma}\cdot{\vec k}\right)\psi({\vec k},t)&+&
\int^{t}_{0}dt'\;  \sigma({\vec k},t-t')\; \dot\psi({\vec k},t')
-\sigma({\vec k},0)\; \psi({\vec k},t)=0 \;,  \nonumber
\end{eqnarray}
with the initial conditions $\psi({\vec k},0)=\psi_0({\vec k})$ and
$\dot\psi({\vec k},t< 0)=0$.

We are now ready to solve the equation of motion by perturbative
expansion in powers of $e^2$ just as in the case of the gauge mean field.
Let us begin by writing
\begin{eqnarray}
&&\psi({\vec k},t)=\psi^{(0)}({\vec k},t) + e^2\;  \psi^{(1)}({\vec
k},t) +{\cal O}(e^4)\; , \nonumber\\
&&\sigma({\vec k},t-t')=e^2\; \sigma^{(1)}({\vec k},t-t')+{\cal
O}(e^4), \nonumber
\end{eqnarray}
we obtain a hierarchy of equations:
\begin{eqnarray*}
\left(i\gamma_0\frac{\partial}{\partial t}
- {\bbox\gamma}\cdot{\vec k}\right)\psi^{(0)}({\vec k},t)&=&0\; ,\\
\left(i\gamma_0\frac{\partial}{\partial t}
- {\bbox\gamma}\cdot{\vec k}\right)\psi^{(1)}({\vec k},t)&=&
\sigma^{(1)}({\vec k},0)\; \psi^{(0)}({\vec k},t)
-\int^{t}_{0}dt'\; \sigma^{(1)}({\vec k},t-t')\; \dot\psi^{(0)}({\vec
k},t') \;,\\ \vdots\quad\quad&&\quad\quad\vdots
\end{eqnarray*}
These equations can be solved iteratively by starting from the
zeroth-order (free field) solution
$$
\psi^{(0)}({\vec k},t)=\sum_{s}\left[ A_s({\vec k}) u_s({\vec k})\;e^{-ikt}
+B_s({\vec k}) v_s(-{\vec k})\;e^{ikt}\right] \;,
$$
and the retarded Green's function of the unperturbed problem
$$
{\cal G}^{(0)}_{\rm ret}({\vec k},t-t')=\left\{-i\gamma_0\cos[k(t-t')]+
{\bbox\gamma}\cdot{\vec\hat k}\sin[k(t-t')]\right\}\;\theta(t-t').
$$
Here $u_s({\vec k})$ and $v_s({\vec k})$ are free Dirac spinors that satisfy
$$
(\gamma_0-{\bbox\gamma}\cdot{\vec\hat k})\left\{
\begin{array}{l}
u_s({\vec k})\vspace{0.5 ex}\\
v_s(-{\vec k})
\end{array}\right\}=0 \;.
$$
The solution to the first-order equation is found to be given by
$$
\psi^{(1)}({\vec k},t)=\psi^{(1,a)}({\vec k},t)+\psi^{(1,b)}({\vec k},t) \;,
$$
where
\begin{eqnarray}
e^2\psi^{(1,a)}({\vec k},t)&=& -i\gamma^{a}({\vec k})\;  t\;
\sum_s \big[A_s({\vec k}) \;  u_s({\vec k})\;e^{-ikt}
-B_s({\vec k}) \;  v_s(-{\vec k})\;e^{ikt}\big],\label{sol1a}\\
e^2\psi^{(1,b)}({\vec k},t)&=&\frac{i}{\pi}\sum_s \int_{-\infty}^{+\infty}
\frac{d\omega}{\omega-k}\;\gamma^{b}(\omega,{\vec k})
\Bigg\{A_s({\vec k})\; u_s({\vec k})\;e^{-ikt}
\left[t-\frac{1-e^{-i(\omega-k)t}}{i(\omega-k)}\right]\nonumber\\
&&-B_s({\vec k})\; v_s(-{\vec k})\;e^{ikt}
\left[t+\frac{1-e^{i(\omega-k)t}}{i(\omega-k)}\right]\Bigg\} \;,
\label{sol1b}
\end{eqnarray}
with
\begin{eqnarray}
\gamma^{a}({\vec k})&=&
\int^{+\infty}_{-\infty}\frac{d\omega}{\omega}\;
\rho_2(\omega,{\vec k}) =\frac{1}{4}\;
\mbox{Tr}\!\left[\mbox{Re}\Sigma(0,{\vec k})
(\gamma_0-{\bbox\gamma}\cdot{\vec\hat k})\right] \; , \cr \cr
\gamma^{b}(\omega,{\vec k})&=&\frac{\pi k}{\omega}
\;[\rho_1(\omega,{\vec k})-\rho_2(\omega,{\vec k})]
=-\frac{k}{4\omega}\;\mbox{Tr}\! \left[\mbox{
Im}\Sigma(\omega,{\vec k}) (\gamma_0-{\bbox\gamma}\cdot{\vec\hat
k})\right] \;.\label{gammab}
\end{eqnarray}
In Eq.~(\ref{sol1a}) secular terms are explicitly
linear in time and are purely imaginary,
whereas in Eq.~(\ref{sol1b}) secular terms may arise at long times from
the resonant denominators.

From the form of $\gamma^{b}(\omega,{\vec k})$ in  Eq.~(\ref{gammab}),
we recognize that $\gamma^{b}(\omega,{\vec k})$ evaluated at the
fermion mass shell $\omega=k$ is the fermion damping rate computed
in perturbation theory~\cite{lebellac}. It has been shown in the
literature that due to the emission and absorption of soft
quasi-static transverse photons which are only dynamically screened by
Landau damping, the fermion damping rate exhibits infrared divergences
near the mass shell in perturbation theory.

This becomes evident from the following analysis.
For soft photons with $q_0,\;q\ll T$, we can replace
$$
1+n_B(q_0)-n_F(p)\simeq T/q_0,\quad
p\simeq k-q\cos\theta \; ,
$$
where $\cos\theta={\vec\hat k}\cdot{\vec\hat q}$, thus write

\begin{eqnarray}
\gamma^{b}(\omega,{\vec k})&=&\frac{\pi e^2 Tk}{\omega}
\int \frac{d^3 q}{(2\pi)^3}\int_{-q}^{+q} \frac{dq_0}{q_0}
\,\Big[(1-\cos^2\!\theta)\,\beta_T(q_0,q)+\beta_L(q_0,q)\Big]\nonumber\\
&&\times\delta(\omega-k+q\cos\theta-q_0).\label{gammab1}
\end{eqnarray}
Here we have neglected the subleading pole contributions which
corresponding to emission and absorption of on-shell
photons~\cite{blaizotBN}. Recall that for very soft $q\ll eT$ the
function $\beta_T(q_0,q)/q_0$ is strongly peaked at $q_0=0$ [see
Eq.~(\ref{breitwigner}) and Fig.~\ref{fig:softphoton}], and as
$q\rightarrow 0$ it can be approximated by
\begin{eqnarray}
\frac{1}{q_0} \; \beta_T(q_0\ll q,q)
&\rightarrow&\frac{\delta(q_0)}{q^2}\quad\mbox{as}
\quad q\rightarrow 0 \; .\label{singular}
\end{eqnarray}
The infrared divergences near the fermion mass shell become manifest
after substituting Eq.~(\ref{singular}) into Eq.~(\ref{gammab1}). The physical
origin of the behavior of the function $\beta_T(q_0,q)/q_0$
as $q_0\ll q\rightarrow 0$ is  the absence of a magnetic mass.

In order to isolate the singular behavior of $\gamma^b(\omega,{\vec k})$,
we follow the steps in Ref.~\cite{blaizotBN} and write
\begin{equation}
\frac{1}{q_0} \;\beta_T(q_0,q)=\delta(q_0)\left(\frac{1}{q^2}-
\frac{1}{q^2+\omega_{P}^2}\right)+
\frac{1}{q_0}\;\nu_T(q_0,q)\;,\label{betatdecomp}
\end{equation}
where $\omega_{P}=eT/3$ is the plasma frequency and
$\nu_T(q_0,q)$ denotes the regular part of the transverse photon
spectral density.
Substituting Eq.~(\ref{betatdecomp}) into Eq.~(\ref{gammab1}), we can
then separate $\gamma^b(\omega,{\vec k})$ into an infrared singular
part which is logarithmically divergent near the fermion mass shell,
given by
\begin{eqnarray*}
\gamma^{b}_{\rm sing}(\omega,{\vec k})&=&\frac{\pi e^2 Tk}{\omega}
\int\frac{d^3 q}{(2\pi)^3}
\left(\frac{1}{q^2}-\frac{1}{q^2+\omega_{P}^2}\right)
\;\delta(\omega-k+q\cos\theta)\;,
\end{eqnarray*}
and a contribution that remains finite and can be expanded near the
fermion mass shell, given by
\begin{eqnarray*}
\gamma^{b}_{\rm reg}(\omega,{\vec k})&=& \frac{\pi e^2
Tk}{\omega}\int\frac{d^3q}{(2\pi)^3}
\int_{-q}^{+q}\frac{dq_0}{q_0}
\;\big[\beta_L(q_0,q)-\cos^2\!\theta \; \beta_T(q_0,q)\\
&&+\;\nu_T(q_0,q)\big]\;\delta(\omega-k+q\cos\theta-q_0)\;.
\end{eqnarray*}
Using the delta function $\delta(q\cos\theta-q_0)$ to perform the angular
integration yields
\begin{eqnarray}
\gamma^{b}(\omega,{\vec k})&\buildrel{\omega\rightarrow k}\over=&
\alpha T\left(\ln\frac{\omega_{P}}{|\omega-k|}
+{\cal I}\right)+{\cal O}(\omega-k)\;,\label{gammabexpanded}
\end{eqnarray}
where
\begin{eqnarray}
{\cal I}&=&\int_0^\infty dq\; q\int_{-q}^{+q} \frac{dq_0}{q_0}\; \bigg[
\beta_L(q_0,q)-\frac{q_0^2}{q^2}\; \beta_T(q_0,q)
+\;\nu_T(q_0,q)\bigg]\;.\label{I}
\end{eqnarray}
The above double integral has been computed analytically in
Ref.~\cite{blaizotBN} with the result ${\cal I}=\ln 3/2$,
\begin{eqnarray*}
\gamma^b(\omega,{\vec k})&\buildrel{\omega\rightarrow k}\over=& -\alpha
T\left(\ln\frac{|\omega-k|}{\omega_{P}}-\frac{\ln 3}{2}\right)+
{\cal O}(\omega-k)\;.
\end{eqnarray*}
We are now in  position to find the secular terms in
$\psi^{(1)}({\vec k},t)$ that emerge in  the intermediate asymptotic regime.

The imaginary part of the secular terms in $e^2\psi^{(1,a)}({\vec k},t)$
and $e^2\psi^{(1,b)}({\vec k},t)$ combine into a linear secular term given by
$$
S_{I,k}(t)=\mp i\,\alpha\,\delta_k\,t\equiv \mp\frac{i\,t}{4}\;
\mbox{Tr}\!\left[\mbox{Re}\Sigma(\omega,{\vec k})
(\gamma_0-{\bbox\gamma}\cdot{\vec\hat k})\right]\!\!\Big|_{\omega=k}\;.
$$
for the positive (negative) energy spinors respectively,
and no further secular terms arise from the higher order
expansion around the fermion mass shell in Eq.~(\ref{gammabexpanded}).
This purely imaginary linear secular term
is thus identified with a perturbative shift of the oscillation
frequency of the mean field~\cite{boyanrgir} and is determined by a
dispersive integral of the spectral densities $\rho_{1,2}(\omega,{\vec k})$
which is rather difficult to obtain in closed form but a detailed
analysis reveals that $\delta_k$ is finite.

The real secular terms are more involved.
The contribution to $\gamma^b(\omega,{\vec k})$ that is finite as
$\omega \rightarrow k$ leads to a linear secular term,
whereas for the logarithmically divergent contribution as
$\omega\rightarrow k$ the following asymptotic
result~\cite{boyanrgir} becomes useful:
\begin{eqnarray}
&&\int_{-\infty}^{+\infty} \frac{dy}{y^2}(1-\cos yt)\ln|y|
\buildrel{t \to \infty}\over= \pi\; t\;(1-\gamma_E-\ln t) +
{\cal O}\left(t^{-1}\right)\;,\nonumber
\end{eqnarray}
where we have neglected terms that fall off at long times.
Thus we find the real part of the secular terms to be given by
$$
S_{R,k}(t)= -\alpha\,T\,t\left(\ln\omega_{P}t +\gamma_E
-1+\frac{\ln 3}{2}\right)\;.
$$
Gathering the above results, at large times $t\gg 1/\omega_{P}
$ the perturbative solution reads
\begin{eqnarray}
\psi({\vec k},t)&=&\sum_{s}\Big\{[1+S_k(t)]\;A_s({\vec k})\;u_s({\vec
k})\;e^{-ikt}
+\;[1+S_k^{\ast}(t)]\;B_s({\vec k})\; v_s(-{\vec k})\;e^{ikt}\Big\}
\nonumber\\&&+\;\text{nonsecular terms}\;,\label{psiper}
\end{eqnarray}
with
$$
S_k(t) = -\alpha\,t\left[\left(\ln\omega_{P}t
+\gamma_E -1+\frac{\ln 3}{2}\right)T+i\,\delta_k\right]\;.
$$

Obviously, this perturbative solution breaks down at a time scale
$\tau_{\rm\,br}\simeq[\alpha T \ln(1/\alpha)]^{-1}$.
To obtain a uniformly valid solution for large times we now implement a
resummation of the secular terms in the perturbative series via the dynamical
renormalization group~\cite{boyanrgir,boyanrgk}.

This method is implemented by defining the (complex) amplitude
renormalization as follows
$$
\begin{array}{l}
A_s({\vec k})={\cal Z}_k(\tau)\;{\cal A}_s({\vec k},\tau)\;,\vspace{1ex}\\
B_s({\vec k})={\cal Z}_k^{\ast}(\tau)\;{\cal B}_s({\vec k},\tau)\;,
\end{array}
$$
where ${\cal Z}_k(\tau)=1+\alpha z_k^{(1)}(\tau)+{\cal O}(e^4)$.
The renormalization coefficient $\alpha z_k^{(1)}(\tau)$ is chosen to
cancel the secular divergence at a time scale $t=\tau$,
i.e, $\alpha z_k^{(1)}(\tau)=-S_k(\tau)$, thus leading to
\begin{eqnarray*}
\psi({\vec k},t)&=&\sum_{s}\Big\{\big[1+S_k(t)-S_k(\tau)\big]
\;{\cal A}_s({\vec k},\tau)\; u_s({\vec k})\;e^{-ikt}\nonumber\\
&& +\;\big[1+S_k^{\ast}(t)-S_k^{\ast}(\tau)\big]\;
{\cal B}_s({\vec k},\tau)\; v_s(-{\vec k})\;e^{ikt}\Big\}\;,
\end{eqnarray*}
which remains bounded at large times provided that $\tau$ is chosen
arbitrarily close to $t$.
The mean field {\em does not} depend on the arbitrary renormalization
scale $\tau$ and this independence leads to the dynamical renormalization
group equation, which to order ${\cal O}(\alpha)$ is given by
\begin{eqnarray}
&&\left[\frac{d}{d\tau}-\frac{d S_k(\tau)}{d\tau}\right]
{\cal A}_s({\vec k},\tau)=0\;,\nonumber \\
&&\left[\frac{d}{d\tau}- \frac{d S_k^{\ast}(\tau)}{d\tau}
\right]{\cal B}_s({\vec k},\tau)=0\;,\nonumber
\end{eqnarray}
with solutions
\begin{eqnarray}
\left\{
\begin{array}{c}
{\cal A}_s({\vec k},\tau)\vspace{1ex}\\
{\cal B}_s({\vec k},\tau)
\end{array}
\right\}&=&
\left\{
\begin{array}{c}
{\cal A}_s({\vec k},\tau_0)\; e^{-i \alpha\delta_k\tau}\vspace{1ex}\\
{\cal B}_s({\vec k},\tau_0)\; e^{i \alpha\delta_k\tau}
\end{array}\right\}
\;e^{-\alpha T\tau\left[\ln\omega_P\tau+0.12652\ldots\right]}\;,\nonumber
\end{eqnarray}
where we have replaced $\gamma_E -1+\ln 3/2 = 0.12652\ldots$, and
$\tau_0\sim 1/\omega_{P} $ is the time scale such that this
intermediate asymptotic solution is valid.
Hence choosing the renormalization point $\tau$
to coincide with the time $t$, we find the long time behavior
(for $\omega_{P} t\gg 1$) of the fermionic mean field for hard
momentum is given by
\begin{eqnarray}
\psi({\vec k},t)&=&\sum_{s}\Big[
{\cal A}_s({\vec k},\tau_0)\; u_s({\vec k})\;e^{-iE(k)t} +
{\cal B}_s({\vec k},\tau_0)\;  v_s(-{\vec k})
\;e^{iE(k)t}\Big]\; e^{-\alpha Tt
\left[\ln\omega_P t+0.12652\ldots\right]}\;,\label{fermion:drgsol1}
\end{eqnarray}
where $E(k)$ is the pole position shifted by one-loop corrections
$$
E(k)=k+\frac{1}{4}
\mbox{Tr}\!\left[\mbox{Re}\Sigma(\omega,{\vec k})(\gamma_0-
{\bbox\gamma}\cdot{\vec\hat k})\right]\!\Big|_{\omega=k}\;.
$$
Eq.~(\ref{fermion:drgsol1}) reveals a time scale for the relaxation
of the fermionic mean field $\tau_{\rm\,rel}\sim[\alpha T \ln(1/\alpha)]^{-1}$,
which coincides with the time scale at which the perturbative solution (\ref{psiper})  breaks down .
This highlights clearly the nonperturbative nature of relaxation phenomena.

This result coincides with that found in Refs.~\cite{blaizotBN,taka}
via the Bloch-Nordsieck approximation and in scalar quantum
electrodynamics~\cite{boyanrgir} using the dynamical renormalization group.

The main purpose of this section was to
introduce the dynamical renormalization group for fermionic theories.
Furthermore, this is another important and relevant
example of the reliability and consistency of this novel
renormalization group applied to real-time nonequilibrium phenomena.

\subsection{Quantum kinetics of the fermion distribution function}

We now study the quantum kinetic equation for the distribution
function of hard fermions. There has  recently been an
intense activity to obtain a Boltzmann equation for quasiparticles in
gauge theories~\cite{bodeker,manuel,BlaizotBOLT} motivated in part by
the necessity to obtain a consistent description for baryogenesis in
non-abelian theories.
Boltzmann equations with a diagrammatic interpretation were obtained
in~\cite{bodeker,BlaizotBOLT} in which a collision-type
kernel describes the scattering of hard quasiparticles.
In these approaches this collision kernel reveals the infrared
divergences associated with the emission and absorption of magnetic
photons (or gluons) and must be cutoff by introducing a relaxation
time scale $\tau_{\rm\,rel} \sim 1/\alpha T$ to leading logarithmic
accuracy~\cite{BlaizotBOLT}.

In a derivation of quantum kinetic equations for charged
quasiparticles in {\em scalar} quantum electrodynamics~\cite{boyanrgir}
using the dynamical renormalization group, it was understood that the origin
of these infrared divergences is the implementation of
Fermi's golden rule that assumes completed collisions and
takes the infinite time limit in the collision kernels.
The dynamical renormalization group leads to quantum
kinetic equations in real time in terms of {\em time-dependent}
scattering kernels without any infrared ambiguity.

In this section we implement this program in spinor QED to derive the
quantum kinetic equation for hard fermions.
There are several important features of our study that must
be emphasized: (i) as presented in detail in Sec.~II, gauge
invariance is automatically taken into account by working
directly with gauge invariant operators, thus the operator that
describes the number of fermionic quasiparticles is gauge invariant,
(ii) a kinetic description relies on a separation
between the microscopic and the relaxation time scales,
this is warranted in a strict perturbative regime and applies to hard
fermionic quasiparticles, (iii) the dynamical renormalization group
leads to a quantum kinetic equation in real time without infrared
divergences since time acts as an infrared cutoff.

The program begins by expanding the Heisenberg fermion field in terms of
creation and annihilation operators as
\begin{eqnarray*}
\psi({\vec x},t)&=&\int\frac{d^3k}{(2\pi)^{3/2}}\; \psi({\vec k},t)\;
e^{i{\vec k}\cdot{\vec x}},\\
\psi^{\dagger}({\vec
x},t)&=&\int\frac{d^3k}{(2\pi)^{3/2}}\;\psi^{\dagger}({\vec k},t) \;
e^{i{\vec k}\cdot{\vec x}}\;,
\end{eqnarray*}
with
\begin{eqnarray*}
\psi({\vec k},t)&=&\sqrt{\frac{m}{\omega_{\vec k}}} \sum_{s}
\left[b_s({\vec k},t)u_s({\vec k})+d^\dagger_s(-{\vec k},t)
v_s(-{\vec k})\right],\\ \psi^\dagger({\vec
k},t)&=&\sqrt{\frac{m}{\omega_{\vec k}}} \sum_{s}
\left[b^\dagger_s(-{\vec k},t)u^\dagger_s(-{\vec k})+d_s({\vec
k},t) v^\dagger_s({\vec k})\right]\;,
\end{eqnarray*}
where $\omega_{\vec k}=\sqrt{{\vec k}^2+m^2}$
and $b_s({\vec k},t)$ [$b^\dagger_s({\vec k},t)$]
is the annihilation (creation) operator that destroys (creates)
a free fermion of momentum ${\vec k}$ and spin $s$ at time $t$.
We have retained the fermion mass to avoid the subtleties
associated with the normalization of massless spinors,
the massless limit will be taken later.

The {\em spin-averaged} number operator for  fermions with momentum
${\vec k}$ is then given by
\begin{eqnarray}
N_f({\vec k},t)&=&\frac{1}{2}\sum_s b^\dagger_s({\vec k},t) \; b_s({\vec
k},t) =\frac{1}{4\omega_{\vec k}} \; \psi^\dagger(-{\vec k},t)(\not\!\!K+m)
 \; \gamma_0 \; \psi({\vec k},t)\;, \nonumber
\end{eqnarray}
where $K=(\omega_{\vec k},{\vec k})$.
Similarly, the {\em spin-averaged} number operator for  anti-fermions
with momentum ${\vec k}$ is then given by
\begin{eqnarray}
\bar{N}_f({\vec k},t)&=&\frac{1}{2}\sum_s d^\dagger_s({\vec
k},t)\;d_s({\vec k},t)
=\frac{1}{4\omega_{\vec k}}\; \psi^\dagger({\vec k},t)(\not\!\!K-m)
\gamma_0\psi({-\vec k},t)\;.\nonumber
\end{eqnarray}
Taking time derivative of $N_f({\vec k},t)$ and using the Heisenberg
equations of motion, we find
\begin{eqnarray*}
\dot{n}^f_{\vec k}(t)&\equiv&\langle\dot{N}_f({\vec k},t)\rangle\\
&=&\lim_{t'\rightarrow t}\frac{ie}{4\omega_{\vec k}}
\int\frac{d^3q}{(2\pi)^{3/2}}
\Big\{\Big\langle\bar{\psi}^-(-{\vec p},t')\big[\gamma_0\; A^-_0(-{\vec q},t')-
\bbox{\gamma}\cdot{\vec A}^{-}_T(-{\vec q},t')\big]\\
&&\;\times(\not\!\!K+m)\;\gamma_0 \; \psi^+({\vec k},t)\Big\rangle\Big\}
+\mbox{c.c.}\;.
\end{eqnarray*}
As before here the ``$+$'' (``$-$'') superscripts for the fields refer
to fields defined in the forward (backward) time branch in the CTP
formulation. It is straightforward to check that
the total fermion number (fermions minus antifermions) is conserved.

In the hard fermion limit $k\sim T\gg m $ we neglect the fermion mass
and obtain to ${\cal O}(e^2)$
\begin{eqnarray}
\dot{n}^f_{\vec k}(t)&=&
e^2\int\frac{d^3q}{(2\pi)^3}\int_{-\infty}^{+\infty}dq_0\int^{t}_{t_0}dt''
\big\{{\cal N}_1(t_0)\cos[(k-p-q_0)(t''-t_0)]\cr \cr
&&\times\;\big[\tilde{\rho}_L(q_0,q) \; {\cal K}^+_{1}({\vec k},{\vec q})
+2\, \tilde{\rho}_T(q_0,q) \;{\cal K}^+_{2}({\vec k},{\vec q})\big]
+\;{\cal N}_2(t_0)\nonumber\\
&&\times\cos[(k+p-q_0)(t''-t_0)]
\big[\tilde{\rho}_L(q_0,q) \; {\cal K}^-_{1}({\vec k},{\vec q})
+2\, \tilde{\rho}_T(q_0,q) \; {\cal K}^-_{2}({\vec k},{\vec q})\big]\big\}\;,
\end{eqnarray}
where ${\cal N}$ and ${\cal K}$ denote respectively the following
statistical and kinematic factors
\begin{eqnarray*}
{\cal N}_{1}(t)&=& [1-n^f_{\vec k}(t)]\;n^f_{\vec
p}(t)\;n_B(q_0)-n^f_{\vec k}(t)\;[1-n^f_{\vec p}(t)][1+n_B(q_0)]\;,\cr \cr
{\cal N}_{2}(t)&=& [1-n^f_{\vec k}(t)][1-n^f_{\vec p}(t)]\;n_B(q_0)
-n^f_{\vec k}(t)\;n^f_{\vec p}(t)\;[1+n_B(q_0)]\; ,\cr \cr
{\cal K}^\pm_{1}({\vec k},{\vec q})&=& 1\pm {\vec\hat k}\cdot{\vec\hat
p} \quad , \quad {\cal K}^\pm_{2}({\vec k},{\vec q})= 1\mp {\vec\hat
k}\cdot{\vec\hat p}\pm
\frac{1-({\vec\hat k}\cdot{\vec\hat q})^2}{1-\frac{q}{k}
({\vec\hat k}\cdot{\vec\hat q})}\;.
\end{eqnarray*}
Here $\tilde{\rho}_T(q_0,q)$ and $\tilde{\rho}_L(q_0,q)$ are the HTL-resummed
spectral densities for the transverse and longitudinal
photons given by Eqs.(\ref{tilderhot}) and (\ref{rholong}), respectively.

In the linearized relaxation time approximation we assume that
$n^f_{\vec k}(t_0)=n_F(k)+\delta n^f_{\vec k}(t_0)$.
Then upon integrating over $t''$, we obtain
\begin{equation}
\delta\dot{n}^f_{\vec k}(t)=-\delta n^f_{\vec k}(t_0)
\int_{-\infty}^{+\infty} d\omega \; {\cal R}_f(\omega,{\vec k}) \;
\frac{\sin[(\omega-k)(t-t_0)]}{\pi(\omega-k)}\;,\label{ndot}
\end{equation}
where
\begin{eqnarray*}
{\cal R}_f(\omega,{\vec k})&=&\pi
e^2\int\frac{d^3q}{(2\pi)^3}\int_{-\infty}^{+\infty}dq_0
\left[1+n_B(q_0)-n_F(p)\right]
\big\{\big[\tilde{\rho}_L(q_0,q) \; {\cal K}^+_{1}({\vec k},{\vec q})\\
&&+\;2\,\tilde{\rho}_T(q_0,q)\;{\cal K}^+_{2}({\vec k},{\vec q})\big]
\;\delta(\omega-p-q_0)+\big[\tilde{\rho}_L(q_0,q)\;{\cal
K}^-_{1}({\vec k},{\vec q})\nonumber\\
&&+\;2\,\tilde{\rho}_T(q_0,q)\;{\cal K}^-_{2}({\vec k},{\vec q})\big]\;
\delta(\omega+p+q_0)\big\}\;.
\end{eqnarray*}
Eq.~(\ref{ndot}) can be integrated directly to yield
\begin{eqnarray}
\delta n^f_{\vec k}(t)&=&\delta n^f_{\vec k}(t_0)\left\{
1-\int_{-\infty}^{+\infty} d\omega\;{\cal R}_f(\omega,{\vec k})
\;\frac{1-\cos[(\omega-k)(t-t_0)]}{\pi(\omega-k)^2}\right\}\;.\label{n1}
\end{eqnarray}
The time-dependent contribution above is now familiar from the
previous discussions, potential secular terms will
emerge at long times from the regions in which the resonant
denominator vanishes. This is the region near the fermion
mass shell $\omega\approx k$, where ${\cal R}_f(\omega,{\vec k})$ is
dominated by the regions of small $q$ and $q_0$, which physically
corresponds to emission and absorption of soft photons. As before for soft
photons with $q,\; q_0 \ll T$, we can replace
\begin{eqnarray*}
&&1+n_B(q_0)-n_F(p)\simeq T/q_0, \quad
p\simeq k-q\cos\theta,\quad
{\cal K}^+_1\simeq 2,\quad
{\cal K}^+_2\simeq 1-\cos^2\!\theta\;,
\end{eqnarray*}
thus write
${\cal R}_f(\omega,{\vec k})$ at $\omega\approx k$ as
\begin{eqnarray}
{\cal R}_f(\omega,{\vec k})&=&2\pi e^2 T \int
\frac{d^3 q}{(2\pi)^3}\int_{-q}^q \frac{dq_0}{q_0}
\;\big[(1-\cos^2\!\theta)\beta_T(q_0,q)+\beta_L(q_0,q)\big]\nonumber\\
&&\;\times\delta(\omega-k+q\cos\theta-q_0)\;.\label{R1}
\end{eqnarray}
Note that the double integral in Eq.~(\ref{R1}) is exactly the same as
that in Eq.~(\ref{I}). Thus ${\cal R}_f(\omega,{\vec k})$ features an
infrared divergence near the fermion mass shell as shown in the
previous subsection.
Following the analysis carried out in the preceding subsection we obtain
\begin{equation}
{\cal R}_f(\omega,{\vec k})\buildrel{\omega\rightarrow k }\over=
-2\alpha T\left[\ln\frac{|\omega-k|}{\omega_{P}}-
\frac{\ln 3}{2}\right]+{\cal O}[(\omega-k)^2]\;.\label{R2}
\end{equation}
Substituting  Eq.~(\ref{R2}) into  Eq.~(\ref{n1}), we find the number of
fermions at intermediate asymptotic times $ t-t_0 \gg 1/\omega_{P} $
to be given by
\begin{eqnarray}
\delta n^f_{\vec k}(t)&=&\delta n^f_{\vec k}(t_0)\{1-2\alpha T(t-t_0)
[\ln\omega_{P}(t-t_0) + 0.12652\ldots]\}\nonumber\\
&&+\;\text{nonsecular terms}.
\end{eqnarray}
As in the case of the fermion mean field relaxation [cf. Eq.~(\ref{psiper})],
the perturbative solution contains a secular term of the form $t\ln t$.
Obviously, the secular term will invalidate the perturbative solution at time
scales $\tau_{\rm\,rel} \simeq[2\alpha T\ln(1/\alpha)]^{-1}$. In the
intermediate asymptotic regime $1/k \ll t-t_0 \ll \tau_{\rm\,rel}$, the
perturbative expansion can be improved by
absorbing the contribution of the secular term at a time scale $\tau$
into a re-definition of the distribution
function. Hence we apply the dynamical renormalization group method
through a renormalization of the
distribution function much in the same manner as the renormalization
of the amplitude in the mean field discussed above,
$$
\delta n^f_{\vec k}(t_0)={\cal Z}_k(\tau,t_0)\; \delta n^f_{\vec k}(\tau)\;,
$$
with
$$
{\cal Z}_k(\tau,t_0)=1+\alpha\;  z_k^{(1)}(\tau,t_0)+{\cal O}(\alpha^2)\;.
$$
The independence of the solution on the time scale $\tau$ leads to the
dynamical renormalization group equation which to lowest order
in $\alpha$ is  given by
\begin{equation}
\left\{\frac{d}{d\tau}+2\alpha
T[1+\ln\omega_{P}(\tau-t_0)+0.12652\ldots ]\right\}
\delta n^f_{\vec k}(\tau)=0\;.\label{fermion:drg}
\end{equation}
\noindent which  is obviously of the form of  a {\em quantum kinetic
equation} in the linearized relaxation time approximation, but with
a {\em time-dependent relaxation rate}.

Solving Eq.~(\ref{fermion:drg}) and choosing $\tau$ to coincide with
$t$, we obtain
the evolution of the fermion distribution function at asymptotic times
$ t-t_0 \gg 1/k$ to be given by
\begin{eqnarray}
\delta n^f_{\vec k}(t)&=&\delta n^f_{\vec k}(t_0) \;
\exp\{-2\alpha T(t-t_0)[\ln\omega_{P}(t-t_0) +\;0.12652\ldots]\}\;.
\label{fermion:drgsol2}
\end{eqnarray}
Comparing Eq.~(\ref{fermion:drgsol2}) with Eq.~(\ref{fermion:drgsol1})
we find that the anomalous exponent that describes the  relaxation of
the fermion distribution function in the linear
approximation is twice that for the linear relaxation of the mean
field. A similar relation is obtained between the
damping rate for the single quasiparticle relaxation and the
relaxation rate of the distribution function in the
case of time-independent rates and true exponential
relaxation~\cite{lebellac}. The dynamical renormalization group reveals
this to be a generic feature even with time-dependent rates.

\section{Conclusions and further questions}

The goals of this article are the study of nonequilibrium effects in
hot QED plasmas directly in real time. The focus is a systematic
study of relaxation of mean fields as well as the distribution
functions for photons and fermions. In particular
the application of the dynamical renormalization group method to study
anomalous relaxation as a consequence of the exchange of soft photons.
To begin with, we have cast our study solely in terms of gauge
invariant quantities, this can be done in the abelian
theory in a straightforward manner, avoiding potential
ambiguities associated with gauge invariance.
The relaxation of photon mean fields revealed important
features: for soft momentum $k\lesssim eT$ mean fields
that are prepared by adiabatically switching-on an
external source there is exponential relaxation towards the
oscillatory behavior dominated by the transverse photon pole.
The source that induces the mean field in this case has a
Fourier transform that is singular at zero frequency and excites the
resonance in the Landau damping region near zero frequency for
the soft photon mean field.
Sources that have a regular Fourier transform
would not lead to the exponential relaxation.
For semihard momentum $eT\ll k \ll T$ in principle both the
HTL and the perturbative approximations are valid,
however the spectral density for photons becomes sharply peaked at the
edge of the Landau damping continuum consistent with
the fact that the photon pole becomes
perturbatively close to the Landau damping cut.
This enhancement of the spectral density near the bare photon
mass shell results in the breakdown of perturbation theory at large times
$kt \gg 1$. The dynamical renormalization
group provides a consistent resummation of the photon self-energy in
real time and leads to anomalous relaxation of the mean field,
given by Eq.~(\ref{finalsol}).
Clearly, higher order terms beyond the HTL approximation will include
collisional contributions leading perhaps to exponential relaxation.
We then expect a crossover between the anomalous power law obtained
in lowest order and the exponential relaxation from collisional processes,
the crossover time scale will depend on the details of the different
contributions and requires a study beyond that presented here.

The dynamical renormalization group provides a consistent and
systematic framework to obtain quantum kinetic equations directly in
real time from the microscopic field theory~\cite{boyanrgir,boyanrgk}.
This method allows to extract information that is {\em not}
available in the usual kinetic description in terms of time-independent
collision kernels obtained under the assumption of completed collisions
which only include on-shell processes.
The dynamical renormalization group approach to quantum kinetics
consistently includes {\em off-shell} processes and accounts for
{\em time-dependent} collisional kernels.
This is important and potentially phenomenologically relevant in
the case of the quark-gluon plasma which has a finite lifetime.

We have implemented this approach to obtain the quantum kinetic
equation for the distribution function of semihard photons in the
case of thermalized fermions in lowest order in the HTL approximation.
This equation features time-dependent rates and we established that
detailed balance, a consequence of fermions being in thermal
equilibrium, emerges on microscopic time scales.
The linearization of the kinetic equation describes relaxation
towards equilibrium with an anomalous exponent, twice as large as that
of the photon mean field in the semihard case.

The kinetic equation for semihard photons allows us to study
photon production from a thermalized quark-gluon plasma by simply
replacing the fermions by two flavors of light quarks.
To leading order in the HTL approximation photon production arises from
{\em off-shell} bremsstrahlung ($q \rightarrow q \gamma$ and
$\bar{q}\rightarrow \bar{q} \gamma$), and the total number of photons
produced at time $t$ is given by Eq.~(\ref{alpha}).
We study the hard region $k\sim T$ numerically and find that for
$T\sim 200\;\mbox{MeV}$ the number of photons produced by these
off-shell processes form a thermalized QGP of lifetime
$10\;\mbox{fm}/c\lesssim t \lesssim 50\;\mbox{fm}/c$
is {\em comparable} to that by on-shell processes estimated in
Refs.~\cite{kapusta,baier} [See Fig.~\ref{fig:photonratio}].

The relaxation of the fermion mean field for hard momentum is
studied with the dynamical renormalization group, we find an
anomalous exponential relaxation which confirms the results
of Refs.~\cite{blaizotBN,taka} where the Bloch-Nordsieck
approximation was used.

We then obtain the quantum kinetic equation for the
distribution function of hard fermions assuming that photons
are in thermal equilibrium. The collisional kernel is
{\em time-dependent} and {\em infrared finite}.
The linearized kinetic equation describes approach
to equilibrium with an anomalous exponential relaxation,
which is twice that of the fermion mean field.

An important payoff of this approach to quantum kinetics is that
it bypasses the assumption of completed collisions which lead to
collisional kernels obtained by Fermi's golden rule and only
describe on-shell processes as in the usual kinetic approach,
which in the case under consideration leads to infrared divergent
collisional kernels~\cite{BlaizotBOLT}.

{\bf Further questions}.
Perhaps the most phenomenologically pressing aspect
that requires further and deeper study is the photon production
by off-shell processes in a thermalized quark-gluon plasma.
This is important in view of the fact that RHIC will begin physics
runs very soon hence an assessment of potential experimental
electromagnetic signatures is very relevant. The next step in the
study of photon production must be to include hard thermal loop
correction to the quark propagators, and to study the spectral
distribution of the photons produced as a function of momentum,
temperature and lifetime of the thermalized QGP.
Furthermore, our calculations just as those in Refs.~\cite{kapusta,baier}
do not account for the expansion of the plasma,
thus an important next step is to couple the quantum
kinetic equation to the hydrodynamic equations of evolution of the QGP,
this is required to obtain the photon distribution function
integrated over the space-time volume of the QGP.
Work on these aspects is currently underway.

Another important extension of this work is to obtain the relaxation
of non-abelian gauge mean fields beyond the HTL approximation
both in the soft and semihard regime. Similarly for the fermionic mean
fields as well as the kinetics of the fermion distribution function.
Of particular importance in this regard would be to include
the nonequilibrium evolution of the distribution function of the gauge
fields in the collision kernels for the fermionic distribution function.
Obviously this will require a consistent analysis of the relevant time scales.
These aspects bear on the physics of baryogenesis and we expect to report
on some of our studies on this issues soon.

\acknowledgements
D.B., H.J.d.V.\ and S.-Y.W.\ would like to thank U.\ Heinz, E.\
Mottola, D.\ Schiff, M.\ Simionato and L.\ Yaffe for useful discussions.
They also thank the Institute for Nuclear Theory at the University of Washington
for its hospitality during the early stages of this work.
D.B. and S.-Y.W.\ thank the US NSF for partial support through grants
PHY-9605186, CNRS-NSF-INT-9815064 and INT-9905954.
S.-Y.W.\ thanks Andrew Mellon Foundation for partial support.
D.-S.L.\ was supported by the ROC NSC through grant NSC89-2112-M-259-008-Y.
LPTHE is UMR 7589 associated to CNRS.

\appendix

\section*{Real-Time Propagators}

In this appendix we summarize the various
CTP real-time propagators used in this article.
\begin{itemize}
\item[(i)]{
The free fermion propagators (with zero chemical potential) are
defined by
\begin{eqnarray}
&&\langle \Psi^{a}({\vec x},t) \bar{\Psi}^{b}({\vec x}',t')\rangle
= i \int \frac{d^3k}{(2\pi)^3} S_{\vec k}^{\,ab}(t,t')\, e^{i{\vec
k}\cdot({\vec x}-{\vec x}')},\nonumber\\ &&S_{\vec k}^{++}(t,t')=
S_{\vec k}^{>}(t,t')\theta(t-t') +S_{\vec
k}^{<}(t,t')\theta(t'-t), \nonumber\\ &&S_{\vec k}^{--}(t,t')=
S_{\vec k}^{>}(t,t')\theta(t'-t) +S_{\vec
k}^{<}(t,t')\theta(t-t'),\nonumber\\ &&S_{\vec k}^{\pm\mp}(t,t')=
S_{\vec k}^{\mbox{\scriptsize
\raisebox{1.8pt}{\raisebox{1.8pt}{$\scriptscriptstyle<$}
\raisebox{-1.5pt}{$\scriptscriptstyle\!\!\!\!\!\!\!\!\;>$}}}}(t,t'),
\label{fermionprop1}
\end{eqnarray}
where $a,\;b=\pm$ and the fermion Wightman functions read
\begin{eqnarray}
S_{\vec k}^{>}(t,t')&=&-\frac{i}{2\omega_{\vec k}}\Big\{
(\not\!{K}+m) [1-n_F(k)] e^{-i\omega_{\vec k}(t-t')}\nonumber\\
&&+\; \gamma_0(\not\!{K}-m)\gamma_0\, n_F(k)\,
e^{i\omega_{\vec k}(t-t')}\Big\},\nonumber \\
S_{\vec k}^{<}(t,t')&=&\frac{i}{2\omega_{\vec k}}
\Big\{(\not\!{K}+m)\,n_F(k)\,e^{-i\omega_{\vec k}(t-t')}\nonumber\\
&&+\;\gamma_0(\not\!{K}-m)\gamma_0 [1-n_F(k)]e^{i\omega_{\vec k}(t-t')}\Big\},
\label{fermionprop2}
\end{eqnarray}
with $K=(\omega_{\vec k},{\vec k})$,
$\omega_{\vec k}= \sqrt{{\vec k}^2+m^2}$ and $n_F(k)$ the
Fermi-Dirac distribution.}

\item[(ii)]{The free (transverse) photon propagators are defined by
\begin{eqnarray}
&&\langle A^{i,a}_T({\vec x},t) A^{j,b}_T({\vec x}',t')\rangle
=-i \int \frac{d^3q}{(2\pi)^3}{\cal G}_{T,q}^{ab}(t,t')\,
{\cal P}_T^{ij}({\vec q})\,e^{i{\vec q}\cdot({\vec x}-{\vec x}')},\nonumber\\
&&{\cal G}_{T,q}^{++}(t,t')=
{\cal G}_{T,q}^{>}(t,t')\theta(t-t')
+{\cal G}_{T,q}^{<}(t,t')\theta(t'-t),\nonumber \\
&&{\cal G}_{T,q}^{--}(t,t')=
{\cal G}_{T,q}^{>}(t,t')\theta(t'-t)
+{\cal G}_{T,q}^{<}(t,t')\theta(t-t'),\nonumber\\
&&{\cal G}_{T,q}^{\pm\mp}(t,t')=
{\cal G}_{T,q}^{\mbox{\scriptsize
\raisebox{1.8pt}{\raisebox{1.8pt}{$\scriptscriptstyle<$}
\raisebox{-1.5pt}{$\scriptscriptstyle\!\!\!\!\!\!\!\!\;>$}}}}(t,t'),
\label{gaugeprop1}
\end{eqnarray}
where
${\cal P}_T^{ij}({\vec q})=\delta^{ij}-\hat{q}^i \hat{q}^j$ is the
transverse projector and
the photon Wightman functions read
\begin{eqnarray}
{\cal G}_{T,q}^{>}(t,t^{\prime})&=&
\frac{i}{2q}\left[[1+n_B(q)]\,e^{-iq(t-t')}+n_B(q)\,
e^{iq(t-t')}\right],\nonumber\\
{\cal G}_{T,q}^{<}(t,t^{\prime})&=&
\frac{i}{2q}\left[n_B(q)\,e^{-iq(t-t')}
+[1+n_B(q)]\,e^{iq(t-t')}\right],\label{gaugeprop2}
\end{eqnarray}
with $n_B(q)$ the Bose-Einstein distribution.}

\item[(iii)]{The HTL-resummed transverse photon propagators are
obtained from the free ones (\ref{gaugeprop1}) by using the
following Wightman functions
\begin{eqnarray}
{\cal G}_{T,q}^{>}(t,t')&=&i\int dq_0
\,\tilde{\rho}_T(q_0,q)\,[1+n_B(q_0)]\,e^{-iq_0(t-t')},\nonumber\\
{\cal G}_{T,q}^{<}(t,t')&=&i\int dq_0\,
\tilde{\rho}_T(q_0,q)\, n_B(q_0)\,e^{-iq_0(t-t')},
\label{wightmanT}
\end{eqnarray}}
where $\tilde{\rho}_T$ is the resumed spectral density.

\item[(iv)]{The HTL-resummed longitudinal photon propagators are
given by~\cite{boyanrgk}
\begin{eqnarray}
&&\langle A^{a}_0({\vec x},t) A^{b}_0({\vec x}',t')\rangle
= i \int \frac{d^3q}{(2\pi)^3}\,{\cal G}_{L,q}^{ab}(t,t')
e^{i{\vec q}\cdot({\vec x}-{\vec x}')},\nonumber\\
&&{\cal G}_{L,q}^{++}(t,t')= \frac{1}{q^2}\delta(t-t')+
{\cal G}_{L,q}^{>}(t,t')\theta(t-t')
+{\cal G}_{L,q}^{<}(t,t')\theta(t'-t),\nonumber \\
&&{\cal G}_{L,q}^{--}(t,t')= -\frac{1}{q^2}\delta(t-t')+
{\cal G}_{L,q}^{>}(t,t')\theta(t'-t)
+{\cal G}_{L,q}^{<}(t,t')\theta(t-t'),\nonumber \\
&&{\cal G}_{L,q}^{\pm\mp}(t,t')=
{\cal G}_{L,q}^{\mbox{\scriptsize
\raisebox{1.8pt}{\raisebox{1.8pt}{$\scriptscriptstyle<$}
\raisebox{-1.5pt}{$\scriptscriptstyle\!\!\!\!\!\!\!\!\;>$}}}}(t,t'),
\end{eqnarray}
with the Wightman functions expressed in terms of the resummed
spectral density $\tilde{\rho}_L$ as
\begin{eqnarray}
{\cal G}_{L,q}^{>}(t,t')&=& -i\int dq_0
\,\tilde{\rho}_L(q_0,q)\,[1+n_B(q_0)]\,e^{-iq_0(t-t^\prime)},\nonumber\\
{\cal G}_{L,q}^{<}(t,t')&=& -i\int dq_0\,
\tilde{\rho}_L(q_0,q)\,n_B(q_0)\,e^{-iq_0(t-t^\prime)}.
\label{wightmanL}
\end{eqnarray}}
\end{itemize}

The free real-time propagators used in studying quantum kinetics are
obtained from (\ref{fermionprop1})-(\ref{gaugeprop2}) with the
equilibrium Fermi-Dirac and Bose-Einstein distributions replacing by
the initial nonequilibrium ones.
Finally, we note that the HTL-resummed photon propagators are only valid
for photons in thermal equilibrium because in deriving their spectral
representations (\ref{wightmanT}) and (\ref{wightmanL})
used has be made of the KMS condition~\cite{boyanrgir,boyanrgk}.




\begin{thebibliography}{99}
\bibitem{qgp1}
J. W. Harris and B. Muller, Annu. Rev. Nucl. Part. Sci. {\bf  46}, 71 (1996);
B. Muller in {\it Particle Production in Highly Excited Matter},
edited by H. H. Gutbrod and J. Rafelski, NATO ASI series B, Vol. 303 (Plenum, New York, 1993);
H. Elze and U. Heinz, Phys. Rep. 183, 81 (1989);
H. Meyer-Ortmanns, Rev. Mod. Phys. {\bf 68}, 473 (1996);
H. Satz, in  {\it Particle Production in Highly Excited Matter},
edited by H.H. Gutbrod and J. Rafelski, NATO ASI series B, Vol. 303 (Plenum, New York, 1993).

\bibitem{qgp2}
B. Muller, {\it The Physics of the Quark-Gluon Plasma},
Lecture Notes in Physics, Vol.~225 (Springer-Verlag, Berlin, 1985);
L. P. Csernai, {\it Introduction to Relativistic Heavy Ion Collisions}
(John Wiley and Sons, England, 1994);
C. Y. Wong, {\it Introduction to High-Energy Heavy Ion Collisions}
(World Scientific, Singapore, 1994).

\bibitem{cern}
For a summary of recent results from SPS, see the CERN web page at
http://www.cern.ch/CERN/Announcements/2000/NewStateMatter/;
U. Heinz and M. Jacob, {\em Evidence for a New State
of Matter: An Assessment of the Results from the CERN Lead Beam
Programme}, nucl-th/0002042.

\bibitem{geiger} K. Geiger, Phys. Rep. {\bf 258}, 237 (1995);
Phys. Rev. {\bf D52}, 1500 (1995);  Phys.Rev. {\bf D54}, 949
(1995); {\em Ultrarelativistic nuclear collisions in a QCD based
space-time description: the parton cascade model} in {\em Quark
Gluon Plasma 2}, Ed. by R. C. Hwa, (World Scientific, Singapore,
1995).

\bibitem{thermalization} K. J. Eskola and X.-N. Wang, Phys. Rev. {\bf D49} 1284 (1994),
K. J. Eskola, B. M\"uller and X.-N. Wang, Phys. Lett. {\bf B374}, 20 (1996);
T. S. Biro et. al., Phys. Rev. C {\bf 48}, 1275 (1993);
T. S. Biro, B. M\"uller and X.-N. Wang, Phys. Lett. B283, 171, (1992).

\bibitem{htl}
E. Braaten and R.D. Pisarski, Nucl. Phys. {\bf B337}, 569 (1990);
{\em ibid}. {\bf B339}, 310 (1990).

\bibitem{rob2}
R.D. Pisarski, Physica A {\bf 158}, 146 (1989);
Phys. Rev. Lett. {\bf 63}, 1129 (1989);
Nucl. Phys. {\bf A525}, 175 (1991).

\bibitem{rob3}
R.D. Pisarski, Nucl. Phys. {\bf B309}, 476 (1988).

\bibitem{weldon1}
H. A. Weldon, Phys. Rev. {\bf D 26}, 1394 (1982);
{\em ibid}. 2789 (1982); Phys. Rev. {\bf D 40}, 2410 (1989);
Physica {\bf A 158}, 169 (1989).

\bibitem{robinfra}
R.D. Pisarski, Phys. Rev. Lett. {\bf 63}, 1129 (1989);
Phys. Rev. D {\bf 47},5589 (1993).

\bibitem{lebellac}
M. Le Bellac, {\em Thermal Field Theory}
(Cambridge University Press, 1996).

\bibitem{inmedium}
H. A. Weldon, Phys. Rev. {D28}, 2007 (1983);
Ann. of Phys. (N.Y.)  {\bf 228}, 43 (1993);
J.-P. Blaizot in Proceedings of the Fourth
Summer School and Symposium on Nuclear Physics,
Eds. D. P. Min and M. Rho (World Scientific, Singapore, 1991).

\bibitem{yaffebrown}
L. S. Brown and  L. G. Yaffe,
{\em Effective Field Theory for Highly Ionized Plasmas}, physics/9911055.

\bibitem{plasmas}
J.-P. Blaizot and E. Iancu, Nucl. Phys. {\bf B459} 559 (1996).

\bibitem{noneq}
J. Schwinger, J. Math. Phys. {\bf 2}, 407 (1961),
K. T. Mahanthappa, Phys. Rev. {\bf 126}, 329 (1962);
P. M. Bakshi and K.T. Mahanthappa, J. Math. Phys. {\bf 41}, 12 (1963),
L. V. Keldysh, JETP {\bf 20}, 1018 (1965), L. P. Kadanoff and G. Baym,
{\em Quantum Statistical Mechanics}, (W. A. Benjamin, New York, 1962),
K.-C. Chou, Z.-B. Su, B.-L. Hao, and L. Yu, Phys. Rep. {\bf 118}, 1 (1985).

\bibitem{disip}
D. Boyanovsky, H.J. de Vega and R. Holman, in Proceedings of
the Second Paris Cosmology Colloquium, Observatoire de Paris, 1994,
edited by H.J. de Vega and N. S\'anchez
(World Scientific, Singapore 1995), pp. 127-215;
in {\em Advances in Astrofundamental Physics},
Erice Chalonge School, edited by N. S\'anchez and
A. Zichichi (World Scientific, Singapore 1995);
D. Boyanovsky, H. J. de Vega, R. Holman, D.-S. Lee and A. Singh,
Phys. Rev. D {\bf 51}, 4419 (1995);
D. Boyanovsky, H. J. de Vega, R. Holman and J. Salgado,
Phys. Rev. D {\bf 54}, 7570 (1996).


\bibitem{tadpole}
D. Boyanovsky, H. J. de Vega and R. Holman,
in {\em Current Topics in Astrofundamental Physics},
Vth Erice Chalonge School, edited by N. S\'anchez and A. Zichichi
(World Scientific, Singapore 1996), pp. 183-270;
D. Boyanovsky, H. J. de Vega, R. Holman and D.-S. Lee,
Phys. Rev. D {\bf 52}, 6805 (1995);
D. Boyanovsky, M. D'Attanasio, H. J. de Vega and R. Holman,
Phys. Rev. D {\bf 54}, 1748 (1996).

\bibitem{photdilep}
L. McLerran and T. Toimela,  Phys. Rev. {\bf D31},
545, (1985); P. V. Ruuskanen,
{\em Photons and lepton pairs--the deep probes of the quark-gluon plasma},
in {\em Particle Production in
Highly Excited Matter}, Ed. Gutbrod, (Plenum, N.Y. 1993);
Nucl. Phys. {\bf A544}, 169c (1995).

\bibitem{kapusta}
J. Kapusta, P. Lichard and D. Seibert,
Phys.\ Rev.\ D {\bf 44}, 2774 (1991).

\bibitem{baier}
R. Baier, M. Dirks, K. Redlich and D. Schiff,
Phys. Rev. {\bf D 56}, 2548 (1997);
R. Baier, H. Nakkagawa, A. Niegawa and K. Redlich,
Z.\ Phys.\ C {\bf 53}, 433 (1992);
R. Baier, B. Pire and D. Schiff,
Phys.\ Rev.\ D  {\bf 38}, 2814 (1988).

\bibitem{boyanrgir}
D. Boyanovsky, H. J. de Vega, R. Holman, and M. Simionato,
Phys. Rev. D {\bf 60}, 065003 (1999);
D. Boyanovsky and H. J. de Vega, Phys. Rev. D {\bf 59}, 105019 (1999).

\bibitem{boyanrgk}
D. Boyanovsky, H.J. de Vega, and S.-Y. Wang,
Phys. Rev. D {\bf 61}, 065006 (2000).

\bibitem{boyanhtl}
D. Boyanovsky, H. J. de Vega, R. Holman, S. P. Kumar, and R. D. Pisarski,
Phys. Rev. D{\bf 58}, 125009 (1998).

\bibitem{rajantie}
A. Rajantie and M. Hindmarsh,
Phys.\ Rev.\  D {\bf 60}, 096001 (1999).

\bibitem{blaizotBN}
J.-P. Blaizot and E. Iancu, Phys. Rev. Lett. {\bf 76}, 3080 (1996);
Phys. Rev. D {\bf 55}, 973 (1997); Phys. Rev. D {\bf 56}, 7877 (1997).

\bibitem{taka}
K. Takashiba, Int. J. Mod. Phys {\bf A 11}, 2309 (1996).

\bibitem{bodeker}
D. Bodeker, Nucl.Phys. {\bf B566}, 402 (2000).

\bibitem{manuel}
D. F. Litim and C. Manuel, Nucl. Phys. {\bf B562}, 237 (1999).

\bibitem{BlaizotBOLT}
J.-P. Blaizot and E. Iancu, Nucl. Phys.
{\bf B557}, 183 (1999); Nucl. Phys. {\bf B570}, 326 (2000).

\bibitem{gaugepot}
D. Boyanovsky, D. Brahm, R. Holman and D.-S. Lee,
Phys. Rev. {\bf D54}, 1763 (1996).

\bibitem{plasmino}
S.-Y. Wang, D. Boyanovsky, H. J. de Vega, D.-S. Lee,
and Y. J. Ng, Phys. Rev. D {\bf 61}, 065004 (2000).

\bibitem{boyanfermion}
D. Boyanovsky, H. J. de Vega, D.-S. Lee, Y. J. Ng, and S.-Y. Wang,
Phys. Rev. D {\bf 59}, 105001 (1999).

\bibitem{baacke}
J. Baacke, D. Boyanovsky, and H. J. de Vega, hep-ph/9907337.

\bibitem{DRG}
L.-Y. Chen, N. Goldenfeld and Y. Oono,
Phys. Rev. Lett. {\bf 73}, 1311 (1994); Phys. Rev. E {\bf 54}, 376 (1996);
T. Kunihiro, Prog. Theor. Phys. {\bf 95}, 503 (1995); {\em ibid}. {\bf 97}, 179 (1997);
S.-I. Ei, K. Fujii and T. Kunihiro, hep-th/9905088;
H. J. de Vega and J.\ F.\ J.\ Salgado, Phys. Rev. D {\bf 56} 6524, (1997);
T. Kunihiro, Prog. Theor. Phys. Suppl. {\bf 131}, 459 (1998);
S.-I. Ei, K. Fujii and T. Kunihiro, Ann. of Phys. (N.Y.) {\bf 280}, 236 (2000);
M. Frasca, Phys. Rev. {\bf A 56}, 1549 (1997);
I. L. Egusquiza and M. A. Valle Basagoiti, Phys. Rev. {\bf D57}, 1586 (1998);
C. M. Bender and L. M. A. Bettencourt, Phys. Rev. Lett. {\bf 77},
4114 (1996), Phys. Rev. {\bf D54}, 7710 (1996).

\bibitem{heinz} H.-T. Elze and U. Heinz, in {\em Quark-Gluon Plasma}
edited by R.C. Hwa (World Scientific, Singapore, 1990); H.-T. Elze
and U. Heinz, Phys. Rep. {\bf 183}, 81 (1989).

\bibitem{mrow} S. Mr\'owczy\'nski, in {\em Quark-Gluon Plasma} \cite{heinz};
S. Mr\'owczy\'nski and P. Danielewicz, Nucl. Phys. {\bf B342}, 345
(1990); S. Mr\'owczy\'nski and U. Heinz, Ann. Phys. (N.Y.) {\bf
229}, 1 (1994); S. Mr\'owczy\'nski, hep-ph/9805435.

\bibitem{daniel} P. Danielewicz, Ann. Phys. (N.Y.) {\bf 152}, 239 (1984).

\bibitem{niegawa}
A. Niegawa, Prog. Theor. Phys. 102, 1 (1999), and references therein.

\end{thebibliography}
\end{document}